\documentclass[a4paper,fleqn,usenatbib,useAMS]{mnras}

\usepackage{graphicx}   
\usepackage{amsmath}    
\usepackage{amssymb}    
\usepackage{multicol}   
\usepackage{bm}         
\usepackage{pdflscape}  
\usepackage{rotating}

\def\la{\;
\raise0.3ex\hbox{$<$\kern-0.75em\raise-1.1ex\hbox{$\sim$}}\; }
\def\ga{\;
\raise0.3ex\hbox{$>$\kern-0.75em\raise-1.1ex\hbox{$\sim$}}\; }

\newcommand{\dmm}{$\Delta\mu/\mu$}

\newcommand{\kms}{km~s$^{-1}$}
\newcommand{\ms}{m~s$^{-1}$}
\newcommand{\etal}{{et al.}}

\usepackage[T1]{fontenc}
\usepackage{ae,aecompl}

\title[Hyperfine structure of CH$_3$OH transitions]
{\textit{
Hyperfine structure of the methanol molecule as traced by Class~I methanol masers
}}

\author[I. I. Agafonova \etal ] {
I. I. Agafonova$^{1}$, O. S. Bayandina$^{2}$, Y. Gong$^{3,4}$, 
C. Henkel$^{3,5}$\thanks{E-mail: chenkel@mpifr-bonn.mpg.de}, Kee-Tae Kim$^{6,7}$,
\newauthor
M. G. Kozlov$^{8,9}$, B. Lankhaar$^{10}$, S. A. Levshakov$^{1}$, K. M. Menten$^{3}$, W. Ubachs$^{11}$,
\newauthor
I. E. Val'tts$^{12}$, W. Yang$^{3,13}$
\vspace*{8pt}
\\
$^{1}$Ioffe Institute, 194021 St.~Petersburg, 26 Polytekhnicheskaya str., Russia\\
$^{2}$INAF--Osservatorio Astrofisico di Arcetri, Largo E. Fermi 5, I-50125 Firenze, Italy\\
$^{3}$Max Planck Institut f\"ur Radioastronomie (MPIfR), Auf dem H\"ugel 69, D-53121 Bonn, Germany\\
$^{4}$Purple Mountain Observatory, and Key Laboratory of Radio Astronomy, Chinese Academy of Sciences, 10 Yuanhua Road,\\
Nanjing 210023, People's Republic of China\\
$^{5}$Xinjiang Astronomical Observatory, Chinese Academy of Sciences, 830011 Urumqi, People's Republic of China\\
$^{6}$Korea Astronomy and Space Science Institute, 776 Daedeokdae-ro, Yuseong-gu, Daejeon 34055, Republic of Korea\\
$^{7}$University of Science and Technology, Korea (UST), 217 Gajeong-ro, Yuseong-gu, Daejeon 34113, Republic of Korea\\
$^{8}$Department of Physics, Electrotechnical University `LETI', 197376 St.~Petersburg, Russia\\
$^{9}$Petersburg Nuclear Physics Institute of NRC `Kurchatov Institute', Gatchina, Leningrad District, 188300, Russia\\
$^{10}$Department of Space, Earth and Environment, Onsala Space Observatory,
Chalmers University of Technology, Onsala, Sweden\\
$^{11}$Department of Physics and Astronomy, LaserLaB, Vrije Universiteit De Boelelaan 1081, 1081 HV Amsterdam,
The Netherlands \\
$^{12}$Astro Space Center, P.N. Lebedev Physical Institute of RAS, 84/32 Profsoyuznaya str., Moscow, 117997, Russia\\
$^{13}$School of Astronomy \& Space Science, Nanjing University, 163 Xianlin avenue, Nanjing 210023,
People's Republic of China\\
}
\date{Accepted 2024 July 11. Received 2024 June 28; in original form 2024 May 15 }

\pubyear{2024}

\begin{document}
\label{firstpage}
\pagerange{\pageref{firstpage}--\pageref{lastpage}}
\maketitle

\begin{abstract}
We present results on simultaneous observations of Class~I methanol masers 
at 25, 36, and 44~GHz towards 22 Galactic targets carried out 
with the Effelsberg 100-m telescope. The study investigates 
relations between the hyperfine (HF) structure of the torsion-rotation transitions
in CH$_3$OH and maser activity. By analyzing the radial velocity shifts between 
different maser lines together with the patterns of the HF structure based on 
laboratory measurements and quantum-chemical calculations, we find that in any 
source only one specific HF transition forms the maser emission and that this 
transition changes from source to source. 
The physical conditions leading to this selective behavior are still unclear. 
Using accurate laboratory rest frequencies 
for the 25~GHz transitions, we have refined the centre frequencies for the HF 
multiplets at 36, 44, and 95~GHz: 
$f_{\scriptscriptstyle 36} = (36169.2488\pm 0.0002_{\scriptscriptstyle\rm stat} \pm
0.0004_{\scriptscriptstyle\rm sys})$~MHz.
$f_{\scriptscriptstyle 44} = (44069.4176\pm0.0002_{\scriptscriptstyle\rm stat} \pm    
0.0004_{\scriptscriptstyle\rm sys})$~MHz,
and
$f_{\scriptscriptstyle 95} = (95169.4414\pm 0.0003_{\scriptscriptstyle\rm stat} \pm
0.0004_{\scriptscriptstyle\rm sys})$~MHz.
Comparison with previous observations of 44~GHz 
masers performed 6-10 years ago with a Korean 21-m KVN telescope towards the same targets 
confirms the kinematic stability of Class~I maser line profiles during this time interval
and reveals a systematic radial velocity shift of $0.013\pm0.005$~\kms\ between the 
two telescopes.
\end{abstract}

\begin{keywords}
methods: numerical --
techniques: spectroscopic --
radio lines: ISM --
ISM: molecules -- 
elementary particles  
\end{keywords}

\section{Introduction}
\label{Sec1}

Methanol (CH$_3$OH) masers\footnote{Maser is short for Microwave Amplification
by Stimulated Emission of Radiation.},
widespread in the Milky Way and also found in nearby galaxies, have proven
to be a powerful tool to study physical and chemical conditions in dense molecular clouds of different origins.
There are two types of these masers, Class~I and Class~II (Batrla \etal\ 1987; Menten 1991).
Class~II masers are characterized by variable and complex emission line
profiles, usually observed close to high-mass young stellar objects 
(Menten \etal\ 1992; Minier \etal\ 2001)
and pumped by infrared (IR) radiation
(Sobolev \& Deguchi 1994; Cragg \etal\ 2005; Green \etal\ 2017).

In contrast, Class~I masers arise frequently relatively far ($\sim 1$ pc) from radiating IR sources
(Menten \etal\ 1986)
and are associated with outflows and shock waves in the interstellar gas 
(Plambeck \& Menten 1990; Cyganowski \etal\ 2009;
Leurini \etal\ 2016; Ladeyschikov \etal\ 2020).
The non-equilibrium inverted population of certain levels
is created in this case by collisions with hydrogen molecules (Leurini \etal\ 2016).
Amplification (exponential in non-saturated and linear in saturated maser regimes)  
of the stimulated emission requires high velocity coherence which implies the
absence of large velocity gradients along the beam path through the gas. 
The masering medium is thus stable and quiet.
As a consequence, the observed maser lines are strong and have narrow and simple profiles,
which allow us to determine the line position with high accuracy.
This makes Class~I methanol masers a preferable tool for precise astronomical measurements,
especially in cases where the analysis relies on the comparison of the
radial velocities of different lines. In particular, this applies to the cases
where astronomical observations are used to refine the data of laboratory
spectroscopy: the observed lines are compared with reference line(s) whose rest frequencies are known
with higher precision. Other examples are
the estimation of magnetic fields by the
Zeeman effect (Vlemmings 2008; Sarma \& Momjian, 2009; Lankhaar \etal\ 2018; Momjian \& Sarma 2019; Sarma \& Momjian 2020)
and the probing of the hypothetical variations of the
electron-to-proton mass ratio, $\mu = m_{\rm e}/m_{\rm p}$
(Jansen \etal\ 2011; Levshakov \etal\ 2011; Dapr\`a \etal\ 2017; Vorotyntseva \etal\ 2024).

\begin{table*}
\centering
\caption{Summary of the observed sources,
their Local Standard of Rest velocities, $V_{\rm \scriptscriptstyle LSR}$,
heliocentric, $D$, and Galactocentric, $R$ distances.
Given in parenthesis
are masers' short names used therein and their other names cited in the literature. }
\label{T1}
\begin{tabular}{r l c c r@.l r@.l r@.l c }
\hline\\[-10pt]
\multicolumn{1}{c}{No.} & \multicolumn{1}{c}{Source}
& R.A. (J2000) & Dec. (J2000) & \multicolumn{2}{c}{$V_{\rm \scriptscriptstyle LSR}$}
& \multicolumn{2}{c}{$D$} & \multicolumn{2}{c}{$R$} & Ref.\\
& & (h:m:s) & ($\circ:\prime:\prime\prime$) & \multicolumn{2}{c}{(km~s$^{-1}$)} &
\multicolumn{2}{c}{(kpc)} & \multicolumn{2}{c}{(kpc)} \\
\hline\\[-10pt]

1 & Orion-KL  & 05:35:14.17 & $-$05:22:46.5 & 7&8 & 0&4&8&7 & 1,2,3 \\[-4pt]
  & ${\rm \scriptscriptstyle (OKL)}$ \\[-2pt]
2 & G208.816$-$19.239 & 05:35:27.14 & $-$05:09:52.5 & 11&4 & 0&4&8&7 & 3,4,5 \\[-4pt]
  & ${\rm \scriptscriptstyle (G208, OMC2)}$ \\[-2pt]
3 & G173.719+2.698 & 05:40:53.30 & $+$35:41:46.9 & $-$17&0 & 1&7&10&0 & 5,6 \\[-4pt]
  & ${\rm \scriptscriptstyle (G173, S235)}$ \\[-2pt]
4 & G183.348$-$0.577 & 05:51:11.15 & $+$25:46:16.4 & 10&0 & 2&0&10&3 & 7 \\[-4pt]
  & ${\rm \scriptscriptstyle (G183)}$ \\[-2pt]
5 & BGPS7501 & 06:12:52.90 & $+$18:00:29.0 & 11&0 & 1&6&9&9 & 8,9 \\[-4pt]
  & ${\rm \scriptscriptstyle (B7501, S255)}$ \\[-2pt]
6 & RMS149 & 06:41:10.15 & $+$09:29:33.6 & 7&2 & 0&6&8&9 & 3,7 \\[-4pt]
  & ${\rm \scriptscriptstyle (R149, NGC2264, G203.316+2.055)}$ \\[-2pt]
7 & RMS153 & 06:47:13.36 & $+$00:26:06.5 & 44&6 & 4&7&12&6 & 7 \\[-4pt]
  & ${\rm \scriptscriptstyle (R153, G212.063-0.741)}$ \\[-2pt]
8 & G013.097$-$0.146 & 18:14:36.90 & $-17$:38:47.5 & 43&7 & 3&8&4&7 & 10 \\[-4pt]
  & ${\rm \scriptscriptstyle (G013)}$ \\[-2pt]
9 & BGPS2147 & 18:20:22.00 & $-16$:14:44.0 & 19&0 & 1&6&6&8 \\[-4pt]
  & ${\rm \scriptscriptstyle (B2147, G14.99-0.70)}$ \\[-2pt]
10 & G018.218$-$0.342 & 18:25:21.99 & $-13$:13:28.5 & 45&9 & 12&3&5&1 & 10 \\[-4pt]
   & ${\rm \scriptscriptstyle (G018)}$ \\[-2pt]
11 & RMS2879 & 18:34:20.89 & $-05$:59:42.5 & 41&8 & 3&0&5&8 & 11 \\[-4pt]
   & ${\rm \scriptscriptstyle (R2879, G25.65+1.05)}$ \\[-2pt]
12 & G029.277$-$0.131 & 18:45:13.88 & $-03$:18:43.9 & 60&1 & 3&6&5&5 & 10 \\[-4pt]
   & ${\rm \scriptscriptstyle (G029)}$ \\[-2pt]
13 & BGPS4518 & 18:47:41.30 & $-02$:00:21.0 & 91&6 & 5&2&5&0 & 10 \\[-4pt]
   & ${\rm \scriptscriptstyle (B4518)}$ \\[-2pt]
14 & RMS3659 & 19:43:11.23 & $+$23:44:03.6 & 22&5 & 2&2&7&5 & 8,12\\[-4pt]
   & ${\rm \scriptscriptstyle (R3659, V645 Cyg)}$ \\[-2pt]
15 & RMS3749 & 20:20:30.60 & $+$41:21:26.6 & 8&8 & 1&4&8&2 & 13 \\[-4pt]
   & ${\rm \scriptscriptstyle (R3749, V1318 CygS)}$ \\[-2pt]
16 & BGPS6815 & 20:35:34.20 & $+$42:20:13.0 & 13&7 & 1&3&8&2 & 10 \\[-4pt]
   & ${\rm \scriptscriptstyle (B6815, G81.302+1.052)}$ \\[-2pt]
17 & BGPS6820 & 20:36:58.10 & $+$42:11:41.0  & 16&3 & 1&3&8&2 & 10 \\[-4pt]
   & ${\rm \scriptscriptstyle (B6820, G81.345+0.760)}$ \\[-2pt]
18 & G102.650+15.786  & 20:39:10.00 & $+$68:01:42.0 & 0&8 & 0&25&8&4 & 14 \\[-4pt]
   & ${\rm \scriptscriptstyle (G102, L1157)}$ \\[-2pt]
19 & BGPS6863 & 20:40:28.70 & $+$41:57:14.0  & $-$6&5 & 3&5&8&6 & 10 \\[-4pt]
   & ${\rm \scriptscriptstyle (B6863, G81.549+0.096)}$ \\[-2pt]
20 & RMS3865 & 20:43:28.49 & $+$42:50:01.8 & 10&3 & 1&4&8&3 & 10 \\[-4pt]
   & ${\rm \scriptscriptstyle (R3865, G82.583+0.201)}$ \\[-2pt]
21 & G095.053+3.972 & 21:15:55.63 & $+$54:43:31.0 & $-$85&6 & 9&0&12&8 & 7 \\[-4pt]
   & ${\rm \scriptscriptstyle (G95)}$ \\[-2pt]
22 & G099.982+4.170 & 21:40:42.40  & $+$58:16:10.0 & $-$0&9 & 0&75&8&5 & 15,16\\[-4pt]
   & ${\rm \scriptscriptstyle (G99, IC 1396 N)}$ \\[-2pt]
\hline\\[-8pt]
\multicolumn{11}{l}{\footnotesize {\it References:} $^1$Barrett \etal\ (1971); $^2$Haschick \etal\ (1990);
$^3$Liechti \& Wilson (1996); $^4$Menten \etal\ (1988); }\\[1pt]
\multicolumn{11}{l}{\footnotesize $^5$Kang \etal\ (2016); $^7$Kim \etal\ (2018); $^8$Kim \etal\ (2019);
$^9$Breen \etal\ (2019); $^{10}$Yang \etal\ (2020); }\\[1pt]
\multicolumn{11}{l}{\footnotesize $^{11}$Bayandina \etal\ (2019); $^{12}$Kalenskii \etal\ (1996);
 $^{13}$Bae \etal\ (2011); $^{14}$Kalenskii \etal\ (2010); }\\[1pt]
\multicolumn{11}{l}{\footnotesize $^{15}$Kalenskii \etal\ (1992); $^{16}$Fontani \etal\ (2010).  }
\end{tabular}
\end{table*}

Methanol is a non-rigid molecule with large-amplitude internal rotation of the methyl group
around the CO bond (Hougen \etal\, 1994; Xu \etal\, 2008).
There are two types of this molecule known:
$A$-methanol with parallel proton spins of the hydrogen atoms in the CH$_3$ group,
and $E$-methanol with a proton with anti-parallel spin.
Spin-rotation, spin-spin and spin-torsion couplings lead to the formation of the hyperfine (HF) structure
(Heuvel \& Dymanus, 1973a,b).
For transitions commonly observed as Class~I methanol masers (at 25, 36, 44, and 95 GHz)
the hyperfine splitting between individual HF components is of order $\sim 10$s~kHz or, on the
velocity scale, of hundreds to tens \ms. Thus, if problems under study require 
line position measurements with an accuracy of 10~\ms\ or better, effects of the HF structure
should be taken into account.

To provide detailed information on the methanol HF structure, a set of high-precision laboratory measurements
was carried out: in the range of $1-25$~GHz~-- with two molecular beam spectrometers
(Coudert \etal\ 2015), and in the range $100-500$~GHz~-- also with two Lamb-dip spectrometers
(Belov \etal\ 2016). These experiments made it possible to measure the centres of certain HF multiplets
with very high accuracy (better than 1 kHz), but the HF structure itself was only partially resolved.
Spectral resolution
of Lamb-dip spectrometers~-- about 10~kHz, i.e., comparable to the HF splitting~-- is simply not high enough
to resolve the HF structure.
The molecular beam spectrometers have much better spectral resolution ($\la 1$~kHz),
but they work with supersonic beams, so that the individual HF components merge due to Doppler broadening.
It is important
to note that laboratory rest frequencies for 36, 44, and 95~GHz multiplets are measured with even lower
accuracy (10 to 30 kHz) and the HF structure of these transitions has never been observed directly.

To evaluate the HF structure theoretically, Lankhaar \etal\ (2016) applied
an original approach to define the effective Hamiltonian:
they calculated a large part of couplings {\it ab initio} and estimated
parameters for the remaining couplings by a fit to the laboratory spectra.
The developed model was then used to calculate the Land\'e $g$-factors
needed to convert the observed Zeeman splitting into the magnetic field strength.
It was shown that in every torsion-rotation multiplet there are only
a few `favored' HF components which can produce maser emission and it was assumed that
only one of these components radiates from any source (Lankhaar \etal\ 2018).

Indeed, a single HF component has been detected in most of the
OH masers thanks to the strong~-- units of MHz~-- HF splittings in the hydroxyl molecule
(see, e.g., reviews by Argon \etal\ 2000, or by Crutcher \& Kemball 2019).
For other masering molecules such as water (H$_2$O), ammonia (NH$_3$) or methanol
with more compact HF splittings
it remains so far unclear whether their maser emission includes a single component,
or is a blend of several HF transitions.

Recently, the presence of a single HF component in methanol masers was confirmed by
Levshakov \etal\ (2022, hereafter L22). Comparing the positions of the maser lines
at 44 and 95~GHz in $A$-methanol they found
that the velocity offsets $\Delta V$ between these lines 
cluster into two groups.
Taking into account that both transitions have only two HF `favored' components,
the revealed bimodality can be explained only under the assumption
that in each source only one HF component is masering and these
components are locked in the 44 and 95~GHz masers. This implies
that
if in the 44~GHz line an HF masering component is shifted
to a higher frequency (blueward) relative to the multiplet centre,
then in the 95~GHz line the masering component will also be shifted bluewards.
The separation between the centres of both groups is $\Delta V = 0.022 \pm 0.003$ \kms\
which is very close to $\Delta V = 0.023$ \kms\ calculated by
the quantum-chemical model of methanol in Lankhaar \etal\ (2016, 2018).
This result implies that
the cited model correctly predicts these favored HF components, at least in the specified multiplets, and that
Class~I methanol masers can be used to explore the HF structure of the methanol molecule.

It was also found that the rest frequencies for the multiplet centres at 44 and 95~GHz are not known 
accurately enough.
It is worth to note that the HF model itself does not involve the calculation of the torsion-rotation
frequencies which are treated simply as external parameters.
Corrections to the rest frequencies can be made either through new laboratory measurements or
by comparison with lines whose rest frequencies are known with higher accuracy.

Another important conclusion relates to the formation mechanism of Class~I masers.
The `favored' components are almost identically pumped, but only one of them becomes a maser and
this component changes from source to source. It is obvious that the process is modulated 
in some way by
physical and chemical conditions in the masering medium.
These conditions are still completely unknown and need to be clarified.

In the present work, we continue to study Class~I methanol masers, this time addressing the $E$-methanol molecule.
With the Effelsberg 100-m radio telescope, we observed $E$-methanol masers at 25 and 36~GHz together with
the $A$-methanol maser at 44~GHz towards 22 Galactic targets.
One aim is to investigate the masering HF components in $E$-methanol
and to determine whether they show the same properties as the masering components
in $A$-methanol, i.e., the existence of only a few favored HF transitions
and the presence of only a single masering HF component in any source.

Another aim is to correct the poorly known rest frequencies for the 36, 44, and 95~GHz HF multiplets
using as a reference accurate laboratory measurements of the 25~GHz
transitions. The implications of these new frequencies for the existing limits on \dmm\ variations as well as
possible mechanisms of Class~I maser formation will be considered in forthcoming papers.

\section{Target selection and observations}
\label{Sec2}

To search for suitable Class~I methanol maser sources, we used the Red {\it MSX} Source (RMS)
catalogue\footnote{https://rms.leeds.ac.uk/cgi-bin/public/RMS\_DATABASE.cgi}
observed by Kim \etal\ (2018), and
the Bolocam Galactic Plane Survey (BGPS)
sources\footnote{https://irsa.opac.caltech.edu/data/BOLOCAM\_GPS}
observed by Yang \etal\ (2020).
The surveys were performed at 44 and 95~GHz with
the Korean Very Long Baseline Interferometry Network (KVN) in single-dish (21-m antenna)
telescope mode.
From these catalogues we selected targets with narrow emission lines located towards both
the Galactic centre and anti-centre and distributed across a wide range of Galactocentric distances,
$5 \la R \la 13$~kpc.

The observations with the Effelsberg 100-m radio telescope
took place during the period 2022 January 26, 28--30 (project number 13-21).
The position-switching mode was used  with the backend eXtended bandwidth 
Fast Fourier Transform Spectrometer (XFFTS)
operating at 300 MHz bandwidth and providing 65\,536 ($2^{16}$) channels
for each polarization.
The methanol lines were measured with
the S14mm Double Beam RX ($\sim$25\,GHz) and the
S7mm Double Beam RX ($\sim$ 36 and 44\,GHz)
installed in the secondary focus.
yielding spectra at 25~GHz with an angular resolution of about $40''$ (HPBW)
in two orthogonally oriented linear polarizations.
The HPBWs were $\simeq 30''$ and $\simeq 25''$ at 36~GHz and 44~GHz, respectively.
The resulting channel separations are 0.055 \kms, 0.038 \kms, and 0.031 \kms\ at
25, 36, and 44~GHz, respectively. However, the true velocity resolution is 1.16 times
coarser (Klein \etal\ 2012).
The telescope pointing was
checked every hour by continuum cross scans of nearby continuum
sources, and the pointing accuracy was better than $5''$.
Depending on the object brightness and weather conditions, 4--24 scans, each lasting 2.4 min, were taken
for every object. This yielded in the co-added spectra signal-to-noise ratio of several 10s and up to
several 100s.
Our sample of 22 objects is presented in Table~\ref{T1}.
Column 2 lists maser source names as they are given in the corresponding catalogues along with
their abbreviated names used therein and other names occurring in the literature (references in the
last column).

\begin{table}
\centering
\caption{List of rest frequencies of the methanol transitions
used in the present work.
The uncertainties are shown in parentheses.
}
\label{T2}
\begin{tabular}{c r@.l c }
\hline\\[-10pt]
\multicolumn{1}{c}{Transition} & \multicolumn{2}{c}{$f$ (MHz)} & Ref. \\
\hline\\[-10pt]
$3_2 - 3_1 E$     &  24928&70105(10) & 1\\
$4_2 - 4_1 E$     &  24933&4702(3)   & 1\\
$5_2 - 5_1 E$     &  24959&0789(4)   & 2\\
$6_2 - 6_1 E$     &  25018&1234(4)   & 1\\
$7_2 - 7_1 E$     &  25124&8719(4)   & 2\\
$4_{-1} - 3_0 E$  &  36169&238(11)   & 3\\
$7_0 - 6_1 A^+$   &  44069&430(10)   & 4\\
$8_0 - 7_1 A^+$   &  95169&463(10)   & 5\\
\hline\\[-8pt]
\multicolumn{4}{l}{\footnotesize {\it References:} 1. Present paper; 2. Mehrotra \etal\ (1985);}\\
\multicolumn{4}{l}{\footnotesize 3. Voronkov \etal\ (2014); 4. Pickett \etal\ (1998);}\\
\multicolumn{4}{l}{\footnotesize 5. M\"uller \etal\ (2004).}\\
\multicolumn{4}{l}{\footnotesize }
\end{tabular}
\end{table}

\section{Line profile analysis}
\label{Sec3}

\subsection{Calculation procedure}
\label{SSec3-1}

The data reduction was performed using the GILDAS software's 
CLASS package\footnote{http://www.iram.fr/IRAMFR/GILDAS/}. The subsequent analysis 
of spectral data was performed in several steps.
At the beginning, narrow spectral segments ($\sim~1.5$ MHz) containing the methanol emission lines
were extracted from the observed scans.
For each interval, the baseline was determined and subtracted from the spectrum.
Individual exposures were added up to enhance the signal-to-noise ratio, S/N.
After that, the mean value of the rms uncertainties, $\sigma_{\rm rms}$,
was calculated using spectral intervals not containing emission lines and/or noise spikes.
The noise in the final spectra is often strongly correlated with 
a consequence that the apparent $\sigma_{\rm rms}$ comes out underestimated.
Approximating the noise with the first-order autoregressive model,
we corrected the noise amplitude as $\sigma^{\rm cor}_{\rm rms} = 
\sigma_{\rm rms}/\sqrt(1-a_{\scriptscriptstyle 1}^2)$, 
where $a_{\scriptscriptstyle 1}$ is the common Pearson
correlation coefficient ($a_{\scriptscriptstyle 1} \simeq 0.7$ in our case).
The obtained value of $\sigma^{\rm cor}_{\rm rms}$ was then assigned to the spectrum. 

We note that absolute flux calibration was not performed due to the poor quality of calibration spectra
at 36 and 44~GHz because of bad weather conditions at the time of observations.
All following calculations are performed  with line intensities given in units of
main-beam brightness temperature, $T_{\rm mb}$.
This does not in any way affect the results since for
the aims of the present work only the kinematic characteristics of the line profiles are relevant.

The transformation to the velocity scale occurred with the rest-frame frequencies listed
in Table~\ref{T2}.
The detailed description of our calculation procedure is given in Levshakov \etal\ (2019).
Here we repeat it only shortly.

Every line profile, which is a function of $v$ on the velocity scale, $\phi(v)$,
is fitted to a kinematic model, $y(v)$, represented by a sum of $N$ Gaussian components.
The line centre is defined as a point
where the first order derivative of the profile function is equal to zero, $y'(v) = 0$. 
This point is further considered as 
the line radial velocity, $V_{\scriptscriptstyle\rm LSR}$. 
We note that in the present context the fitting model acts simply as a filter which is
employed to smooth out eventual small-scale fluctuations
which can hamper the evaluation of $y'(v)$.

The parameters of the fitting function are calculated by a standard $\chi^2$ minimization:
\begin{equation}
\chi^2_\eta = \frac{1}{\eta}\sum^n_{i=1} \{[\phi(v_i) - y(v_i)]/\sigma^{\rm cor}_{\rm rms}\}^2,
\label{Eq3-1}
\end{equation}
where $n$ is the number of channels covering the line profile, and
$\eta$ is the number of degrees of freedom, $\eta = n-3N$.

The number of Gaussian components is chosen so that the $\chi^2_\eta$ function
is minimized at the level of $\chi^2_\eta \simeq 1$ to avoid under- or
over-fitting of the line profile.
The uncertainty of $V_{\scriptscriptstyle\rm LSR}$, $\sigma_v$, is determined by
three points $\{v_1,y_1; v_2,y_2; v_3,y_3\}$ with $v_1 < v_2 < v_3$
which include the flux density peak, $v_{\rm peak} \in (v_1,v_3)$:
\begin{equation}
\sigma_v = \frac{\sigma^{\rm cor}_{\rm rms} \Delta_{\rm ch} {\cal K}}{(y_1 - 2y_2 + y_3)^2}\ ,
\label{Eq3-2}
\end{equation}
where ${\cal K} =  \sqrt{(y_3-y_2)^2 + (y_1-y_3)^2 + (y_2-y_1)^2}$, and
the channel width $\Delta_{\rm ch} = v_2-v_1 = v_3-v_2$.

\begin{table*}
\centering
\caption{LSR radial velocities of the $7_0 - 6_1 A^+$ transition at 44 GHz in Class~I methanol masers
measured with the 
KVN in single-dish (21-m antenna) mode
($V_{\scriptscriptstyle \rm KVN}$) and the Effelsberg 100-m telescope
($V_{\scriptscriptstyle \rm Eff}$) at different epochs.
Statistical errors ($1\sigma$) in the last digits are given in parentheses.
}
\label{T3}
\begin{tabular}{l c r@.l c r@.l r@.l }
\hline\\[-10pt]
\multicolumn{1}{c}{Source} & \multicolumn{3}{c}{KVN}
& \multicolumn{3}{c}{Effelsberg}
& \multicolumn{2}{c}{$\Delta V = $} \\
\multicolumn{1}{c}{ID} & Date & \multicolumn{2}{c}{$V_{\scriptscriptstyle \rm KVN}$} &
Date & \multicolumn{2}{c}{$V_{\scriptscriptstyle \rm Eff}$} &
\multicolumn{2}{c}{$V_{\scriptscriptstyle \rm Eff} - V_{\scriptscriptstyle \rm KVN}$}\\
 & & \multicolumn{2}{c}{(km~s$^{-1}$)} & & \multicolumn{2}{c}{(km~s$^{-1}$)} &
\multicolumn{2}{c}{(km~s$^{-1}$)} \\
\hline\\[-10pt]
R153  &    02.2012 &  44&644(6) & 01.2022   &  44&684(2)   &  0&040(6) \\
R149  &    03.2012 &   7&216(7) & 01.2022   &   7&251(2)   &  0&035(7) \\
R2879 &    05.2012 &  41&780(2) & 01.2022   &  41&778(2)   & $-0$&002(3) \\
R3659 &    10.2012 &  22&495(5) & 01.2022   &  22&518(2)   &  0&023(5) \\
R3749 &    10.2012 &  8&826(4)  & 01.2022   &   8&840(4)   &  0&014(5) \\
R3865 &    10.2012 &  10&345(5) & 01.2022   &  10&322(2)   & $-0$&023(5) \\
B7501 &    11.2016 &  11&036(3) & 01.2022   &  11&053(2)   &  0&017(4) \\
B2147 &    11.2016 &  19&033(17) & 01.2022  &  19&048(2)   &  0&015(17) \\
B4518 &    11.2016 &  91&558(9)  & 01.2022  &  91&579(7)   &  0&021(11) \\
B6815(1) & 11.2016 &  13&703(6)  & 01.2022  &  13&708(2)   &  0&005(6) \\
B6815(2) & 11.2016 &  14&073(8)  & 01.2022  &  14&090(6)   &  0&017(10) \\
B6820(1) & 11.2016 &  16&346(5)  & 01.2022  &  16&375(13)  &  0&029(14) \\
B6863  &   11.2016 & $-6$&516(10) & 01.2022 &  $-6$&488(5) &  0&028(11) \\
G029   &   11.2016 &  60&100(5)   & 01.2022 &   60&141(2)  &  0&041(5) \\
G018(1)  & 11.2016 &  45&880(6)   & 01.2022 &   45&882(8)  &  0&002(10)\\
G018(2)  & 11.2016 &  46&498(13)  & 01.2022 &   46&500(18) &  0&002(22)\\[2pt]
\multicolumn{7}{r}{{\sl weighted mean} $\langle \Delta V \rangle$:} & 0&013(5)\\
\hline\\[-8pt]
\end{tabular}
\end{table*}

\begin{table*}
\centering
\caption{Parameters of the HF components
evaluated from the fitting of the methanol laboratory lines shown in Fig.~\ref{F1}.
The line centre, full width at half maximum, and amplitude are labeled as
$V_i$, FWHM$_i$, and ${\cal A}_i$, respectively.  
The frequency offsets, $\Delta_i f$, are given relative to the central
frequency $f_{\scriptscriptstyle 0}$, also given in Table~\ref{T1}. 
Component numbers (1st column) correspond to those depicted in Fig.~\ref{F1}.
Statistical errors ($1\sigma$) in the last digits, calculated via inversion of the Hesse matrix,
are given in parentheses.
}
\label{T4}
\begin{tabular}{c r@.l  r@.l r@.l r@.l  r@.l  r@.l r@.l r@.l }
\hline\\[-10pt]
No. & \multicolumn{2}{c}{$V_i$} & \multicolumn{2}{c}{FWHM$_i$} & \multicolumn{2}{c}{${\cal A}_i$} &
\multicolumn{2}{c}{$\Delta_i f$} &
\multicolumn{2}{c}{$V_i$} & \multicolumn{2}{c}{FWHM$_i$} & \multicolumn{2}{c}{${\cal A}_i$} &
\multicolumn{2}{c}{$\Delta_i f$}\\
 & \multicolumn{2}{c}{(\kms)} & \multicolumn{2}{c}{(\kms)}  &\multicolumn{2}{c}{ }  & \multicolumn{2}{c}{(MHz)} &
\multicolumn{2}{c}{(\kms)} & \multicolumn{2}{c}{(\kms)}  &\multicolumn{2}{c}{ }  & \multicolumn{2}{c}{(MHz)} \\
\hline\\[-10pt]
\multicolumn{9}{c}{\sl $J=3, f_{\scriptscriptstyle 0}=24928.70105(10)$ MHz} &
\multicolumn{8}{c}{\sl $J=5, f_{\scriptscriptstyle 0}=24959.0789(4)$ MHz}\\
 1 & $-0$&29(6)  & 0&49(12)  & 0&60(12) & 0&024(5)    & $-0$&30(4) & 0&19(8) & 0&44(15) & 0&025(3)\\
 2 & $-0$&177(9) & 0&108(17) & 0&42(6)  & 0&0147(7)   & $-0$&252(19) & 0&03(3) & 0&06(6) & 0&0210(16)\\
 3 & $-$0&111(3) & 0&111(5)  & 1&54(6)  & 0&0093(2)   & $-0$&174(9) & 0&092(17) & 0&67(11) & 0&0144(8)\\
 4 & $-0$&043(6) & 0&069(11) & 0&34(5)  & 0&0035(5)   & $-0$&095(6) & 0&094(11) & 1&04(11) & 0&0079(5)\\
 5 &  0&059(4)  & 0&052(7)  & 0&33(4)  & $-0$&0049(4) & 0&001(6) & 0&101(12) & 1&07(11) &$-0$&0001(5) \\
 6 &  0&109(2)  & 0&086(4)  & 1&34(5)  & $-0$&0090(2) & 0&101(6) & 0&107(11) & 1&29(12) &$-0$&0084(5)\\
 7 &  0&175(5)  & 0&094(9)  & 0&62(5)  & $-0$&0145(4) & 0&19(2)  & 0&11(4) & 0&41(12) & $-0$&0157(16) \\
 8 &  0&27(3)   & 0&12(5)   & 0&17(6)  & $-0$&023(2)  & 0&25(7)  & 0&30(12)& 0&6(2)  & $-0$&021(5)\\
\multicolumn{9}{c}{\sl $J=4, f_{\scriptscriptstyle 0}=24933.4702(3)$ MHz} & \multicolumn{8}{c}{\sl $J=6,
f_{\scriptscriptstyle 0}=25018.1234(4)$ MHz}\\
 1 &  $-0$&27(2)   & 0&08(4)   & 0&13(6)  &  0&0228(19) & $-0$&35(3)&  0&05(5) & 0&20(18) & 0&029(2)\\
 2 &  $-0$&194(13) & 0&06(2)   & 0&14(5)  &  0&0161(11) & $-0$&25(2) & 0&07(4) & 0&4(2)  & 0&0206(17)\\
 3 &  $-0$&109(4)  & 0&107(8)  & 1&10(7)  &  0&0091(4)  & $-0$&186(17) &0&04(3) & 0&24(16) & 0&0155(15)\\
 4 &  $-0$&034(8)  & 0&072(16) & 0&32(6)  &  0&0028(7)  & $-0$&120(8) &0&090(16) & 1&5(2) & 0&0100(7)\\
 5 &    0&047(3)   & 0&072(6)  & 0&84(6)  & $-0$&0039(3)& $-0$&043(10) &0&063(17) & 0&77(19) & 0&0036(8) \\
 6 &    0&114(2)   & 0&117(3)  & 2&92(8)  & $-0$&0094(2) & 0&03(2) & 0&04(3) & 0&19(15) & $-0$&0029(16)\\
 7 &    0&207(8)   & 0&046(15) & 0&16(5)  & $-0$&0172(7) & 0&099(9) & 0&076(17) & 1&1(2) & $-0$&0082(8)\\
 8 &    0&27(2)    & 0&11(4)   & 0&23(7)  & $-0$&0223(17) & 0&18(2) &  0&07(4) & 0&4(2) & $-0$&0153(18)\\
\hline\\[-8pt]
\end{tabular}
\end{table*}

\subsection{Reproducibility of radial velocities in Class~I methanol masers at 44 GHz}
\label{SSec3-2}

A number of methanol maser transitions from the present study have been previously observed
using different facilities at different radio telescopes. We can compare the corresponding
data in order to test the kinematic stability of Class~I methanol masers over a 10-yr time lapse
and to estimate possible systematics between telescopes.

Previously, such comparison was performed in L22 for the 44~GHz line in 10 objects observed with
the KVN in single-dish mode in two surveys between epochs 2012 and 2016.
In all objects, no significant shifts of the line centres were detected.

For objects from the present study, Table~\ref{T3} shows $V_{\scriptscriptstyle\rm LSR}$ evaluated for
the 44~GHz lines taken in January, 2022 with the Effelsberg 100-m telescope (HPBW~$\simeq 40''$)
and the 44~GHz lines observed at different epochs with the 
KVN (HPBW~$\simeq 65''$).
The spectral resolution was
0.031 \kms\ and 0.053 \kms\ for the 100-m and 21-m telescopes, respectively.

The last column in Table~\ref{T3} gives the differences between the peak radial velocities,
$\Delta V =  V_{\scriptscriptstyle \rm Eff} - V_{\scriptscriptstyle \rm KVN}$.
The weighted mean value of $\Delta V = 0.013\pm0.005$ \kms\ ($1\sigma$ error of the mean)
can be interpreted as a systematic error inherent to the performed observations.
Different apertures, instrumental setups and observational conditions~-- all these factors
contribute to the specified error.
Taking into account the indicated complexity, we can
consider the revealed systematic error as acceptable.

Another point is the physical stability of methanol masers themselves.
As was already mentioned above, Class~I methanol masers are suggested to
be kinematically stable. In general, our results confirm this.
Their fluxes~-- by definition a
much more volatile characteristic than the velocity field~--  also vary very slowly on the time scale
of a few years which means that they emit to a large extent
in a regime of saturation (Menten \etal\ 1988; Kurtz \etal\ 2004; Yang \etal\ 2020; Wenner \etal\ 2022).
Assuming that these masers are confined to shocks and have characteristic sizes
of $\sim~50$ AU, the time intervals when variability of the line position will
be perceptible can be estimated as $\sim~15$~yr (Leurini \etal\ 2016).
Table~\ref{T3} lists 6 objects observed within the 10-yr time lapse, and just three of them
(R149, R153, R3865) from this set
show statistically significant (over $5\sigma$) velocity
shifts from the mean value, whereas from 10 objects
observed within 6 years the comparable shift demonstrates only one (G029).
Of course, statistics are rather poor and, additionally,
without accurate absolute flux calibration we cannot conclude
definitely whether these outliers are due to some 
shortcomings in the observations/data processing or due to real physical processes. 
This problem will be addressed in future studies.

\subsection{Laboratory measurements of methanol HF rest frequencies at 25 GHz}
\label{SSec3-3}

High-dispersion laboratory spectroscopy
of methanol torsion-rotation transitions at low frequencies
were performed by J.-U.~Grabow and S.~A.~Levshakov with the microwave molecular beam
spectrometer at the Leibniz University Hannover in 2012.
Among others, the following lines
of the $E$-type ground torsional state ($v_t = 0$) of CH$_3$OH were recorded:
$J_2 \to J_1 =$
$3_2-3_1E$ (24928 MHz), $4_2-4_1E$ (24933 MHz), $5_2-5_1E$ (24959 MHz),
and $6_2-6_1E$ (25018 MHz).
The experimental setup was described briefly in Coudert \etal\ (2015). In the observed spectra,
the HF structure was only partially resolved: we saw bi- and trimodal patterns where
each mode consists of several blended lines.
Primary processing of these spectra was aimed at calculating the accurate values of the HF multiplet
centres which are now presented in Table~X in Coudert \etal\ (2015).
The reconstruction of the methanol HF structure was out of the scope of that work.
Recently, the observed morphology of the line shapes at 25~GHz was studied more closely by
Vorotyntseva \& Levshakov (2024).

\begin{table}
\centering
\caption{Velocity splittings between the HF modes of
the partially resolved laboratory profiles of methanol
lines (Fig.~\ref{F1}).
Listed are the mode centres
$V_1$ and  $V_2$, and their differences
$\Delta V = V_2 - V_1$.
The $1\sigma$ uncertainties in the last digits are given in parentheses.
}
\label{T5}
\begin{tabular}{c r@.l r@.l r@.l}
\hline\\[-10pt]
Transition & \multicolumn{2}{c}{$V_1$} & \multicolumn{2}{c}{$V_2$} &
\multicolumn{2}{c}{$\Delta V$}\\
 & \multicolumn{2}{c}{(km~s$^{-1}$)} & \multicolumn{2}{c}{(km~s$^{-1}$)} & \multicolumn{2}{c}{(km~s$^{-1}$)} \\
\hline\\[-10pt]
$3_2-3_1E$ &  $-0$&116(3) &  0&111(3) &   0&227(4)\\
$4_2-4_1E$ &  $-0$&103(5) &  0&095(5) &   0&198(7)\\
$5_2-5_1E$ &  $-0$&097(7) &  0&100(9) &   0&197(11)\\
$6_2-6_1E$ &  $-0$&117(16) & 0&100(4) &   0&217(16)\\
\hline\\[-8pt]
\end{tabular}
\end{table}

Here we re-process the obtained laboratory spectra in 2012 in order to see whether we can reveal
the detailed HF structure of methanol lines in question.
The idea behind this is that the convolved components,
even if not fully resolved, affect nevertheless the shape of the line envelope
and~-- assuming a sufficiently high signal-to-noise ratio~-- can be more
or less accurately deconvolved.
The spectra have a spacing of $\Delta f = 1.2$~kHz ($\Delta v = 0.014$~\kms),
thus making it possible to distinguish components detached by $\sim 10$s of kHz.

The measurements were made in a supersonic molecular beam (CH$_3$OH/Ne mixture)
injected parallel to the axis of the Fabry-P\'erot resonator.
The CH$_3$OH emission was generated by an external excitation impulse of the corresponding frequency.
Ideally, a standing wave is formed in the resonator with nodes exactly on the axis;
this produces an emission signal in the form of a doublet
consisting of two symmetric parts Doppler-shifted relative to the central node (resonance frequency)
which will be referred to as image~I and image~II hereinafter.
However, in the case of non-resonant excitation
of the molecular emission and/or non-tuned resonator the resulting
spectral profiles can come out distorted (Grabow 2004, 2011).
It is clear that parameters extracted from such profiles also can be biased.

Next, the Hannover University spectrometer was affected
by residual magnetic fields of $B \la 0.5$ Gauss~-- this may cause shifts in the component centres
of about units of kHz.
Special care is also required when measurements are carried out in the frequency range
approaching the instrumental limit, which for the Hannover spectrometer is 26.5~GHz.
In this case, the signal becomes weak and must be amplified, so that the general noise
level increases even more due to the addition of the amplifier noise.
It is obvious that the accuracy of the extracted parameters will be negatively impacted.
Thus, both the raw data and the results of calculations should be treated with caution.

\begin{figure*}
\vspace{-1cm}
\centering
\includegraphics[width=1.00\textwidth]{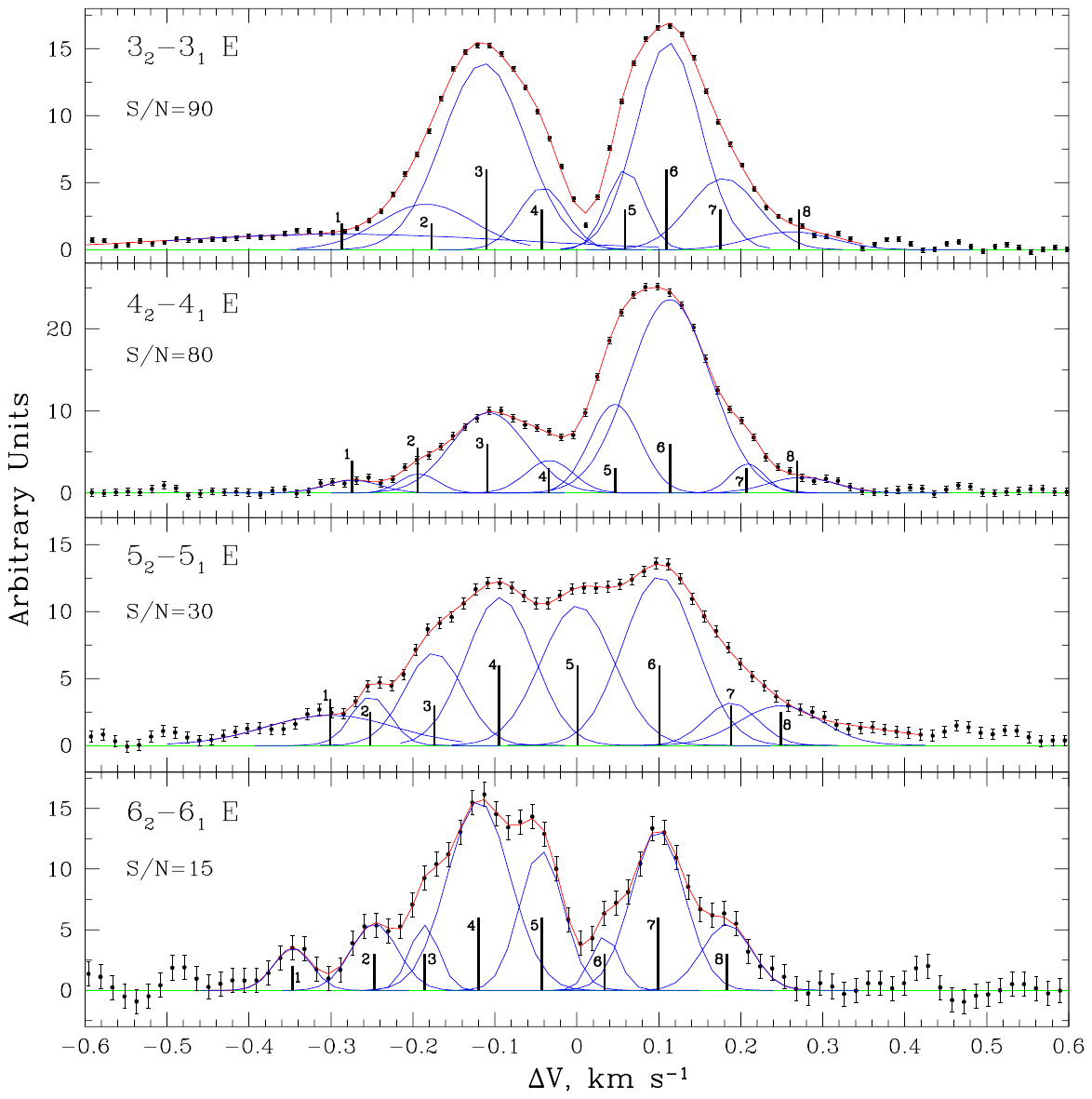}
\vspace{-4.0cm}
\caption{\small
Dots with error bars display the laboratory spectra of methanol transitions at 25~GHz
recorded with the molecular beam spectrometer of the Hannover University.
The individual components are the blue curves, whereas their convolution is shown in red.
The component fitting parameters are listed in Table~\ref{T4}.
}
\label{F1}
\end{figure*}

The experimental profiles of the 25~GHz lines are shown in Fig.~\ref{F1} by dots with error bars representing
the rms of the noise which was calculated using emission-free parts of the spectra on the
left- and right-side to the emission line.
The multiplet central frequencies listed in Table~\ref{T2}
were obtained by cross-correlating both images of the signal.
Then these frequencies were used to transform the observed spectra to the velocity scale.

\begin{table}
\centering
\caption{Source RMS3865: fitting parameters for the thermal methanol emission lines
at 25~GHz (Fig.~\ref{F2}).
Listed are the HF mode centres $V_1$ and $V_2$,
splitting between the modes $\Delta V = V_2-V_1$,
and the total line width FWHM.
The $1\sigma$ uncertainties in the last digits are given in parentheses.
}
\label{T6}
\begin{tabular}{l c r@.l r@.l c }
\hline\\[-10pt]
\multicolumn{1}{c}{Transition} & $V_1$ & \multicolumn{2}{c}{$V_2$} &
\multicolumn{2}{c}{$\Delta V$} & FWHM \\
 & \multicolumn{1}{c}{(km~s$^{-1}$)} & \multicolumn{2}{c}{(km~s$^{-1}$)} & \multicolumn{2}{c}{(km~s$^{-1}$)} &
(km~s$^{-1}$) \\
\hline\\[-10pt]
$4_2-4_1E$  &  10.46(4)  & 10&76(10)   & 0&30(11)  & 0.5(8)\\
$5_2-5_1E$  &  10.45(3)  & 10&66(3)    & 0&21(4)   & 0.5(5)\\
$6_2-6_1E$  &  10.46(2)  & 10&66(2)    & 0&21(3)   & 0.5(3)\\
$7_2-7_1E$  &  10.46(3)  & 10&69(3)    & 0&23(4)   & 0.5(5)\\
\hline\\[-8pt]
\end{tabular}
\end{table}

Below some comments are given on the spectra pictured in Fig.~\ref{F1}:
\begin{itemize}
\item Line $3_2-3_1E$~-- the second image of this line
is slightly distorted and was discarded, only one image was used
in our analysis.
However, the S/N ratio is high enough (S/N = 90) to provide an accurate
deconvolution of the subcomponents even from a single image.
\item Line $4_2-4_1E$~-- both images~I and II are similar and were co-added to produce
the resulting spectrum with S/N = 80.
However, here the amplitude of one mode was noticeably lower than that of the other mode
whereas in other lines both modes look more or less identical.
This can hint at possible experimental flaws in the $4_2-4_1E$ transition measurement.
As a consequence, the derived parameters cannot be considered reliable despite small statistical errors,
i.e., the revealed HF structure requires confirmations by further measurements/observations.
\item Line $5_2-5_1E$~-- both images were affected by a high noise.
To reduce the noise,
individual images were processed with moving average filters with a three-point window
and then co-added.
\item Line $6_2-6_1E$~-- the images were filtered in the same way as for the $5_2-5_1E$ line.
Apart from general noise, one of the images
revealed also an extended outlying fragment which could not be filtered out.
This fragment was removed from the profile and the
remaining parts were co-added with the second image.
\end{itemize}

The obtained line profiles were fitted to a model represented by a sum of $N$ Gaussian components, 
each of which is
characterized by three parameters~-- the component centre ($f_i$), the full width at half maximum (FWHM$_i$),
and the amplitude ($A_i$).
The parameters were estimated by a standard $\chi^2$ minimization.
The value of $N$ was chosen by iterations as the  {\it minimum} number of components for which the
condition $\chi^2_\eta \simeq 1$ is fulfilled.

The results of the calculations are presented in Table~\ref{T4} with individual components pictured
in Fig.~\ref{F1} in blue.
The parameter errors given in Table~\ref{T4} are purely statistical
(obtained by inversion of the Hesse matrix).
However, due to the reasons mentioned above (line distortion, residual magnetic fields)
there could also be systematic errors of similar values.
We note that the Gaussian
subcomponents have different FWHMs ranging from  FWHM $\sim~0.05$ \kms\ to $\sim~0.5$ \kms.
Most likely, the narrowest components
represent single HF lines, whereas components with larger dispersions are blends of close
HF components which cannot be resolved with the current S/N and spacing in the data available.

Another result worth noting is the presence of a strong component near
the multiplet centre in the $5_2-5_1E$ and $6_2-6_1E$ lines (component \#5
in Table~\ref{T4} and in Fig.~\ref{F1}).
Such a component is not seen in the $3_2-3_1E$ spectrum.
As for the $4_2-4_1E$ line, since its profile is presumably distorted,
the conclusions about the presence of a strong central component require
additional laboratory measurements and/or astronomical observations.

Given in Table~\ref{T5} are the splittings between two convolved HF modes.
It is a useful parameter which can be utilized to test the quantum-chemical models
(Coudert \etal\ 2015; Belov \etal\ 2016; Lankhaar \etal\ 2016; Vorotyntseva \& Levshakov 2024),
or to conclude whether the double-peak profile is produced by a chance overlapping of
two separate lines or indeed by convolved HF components of the same torsion-rotation transition.
It is seen that in our case the splitting
does not demonstrate a dependence on the rotational angular momentum $J$~--
unlike the result reported for high $J$ in Belov \etal\ (2016).
Further on, comparing line profiles calculated on base of Lankhaar's methanol model
(Figs.~2 and 3 in Lankhaar \etal\, 2016) with our laboratory profiles and their deconvolution, we see
that for the $3_2-3_1E$, $4_2-4_1E$ and $5_2-5_1E$ transitions near 25~GHz this model
yields an unsatisfactory result and should be refined.
In contrast to this we recall here that for
the 44 and 95~GHz transitions in $A$-methanol the model was very successful.

\subsection{Astronomical observation of the methanol HF structure at 25 GHz}
\label{SSec3-4}

\begin{figure*}
\centering
\includegraphics[width=0.70\textwidth]{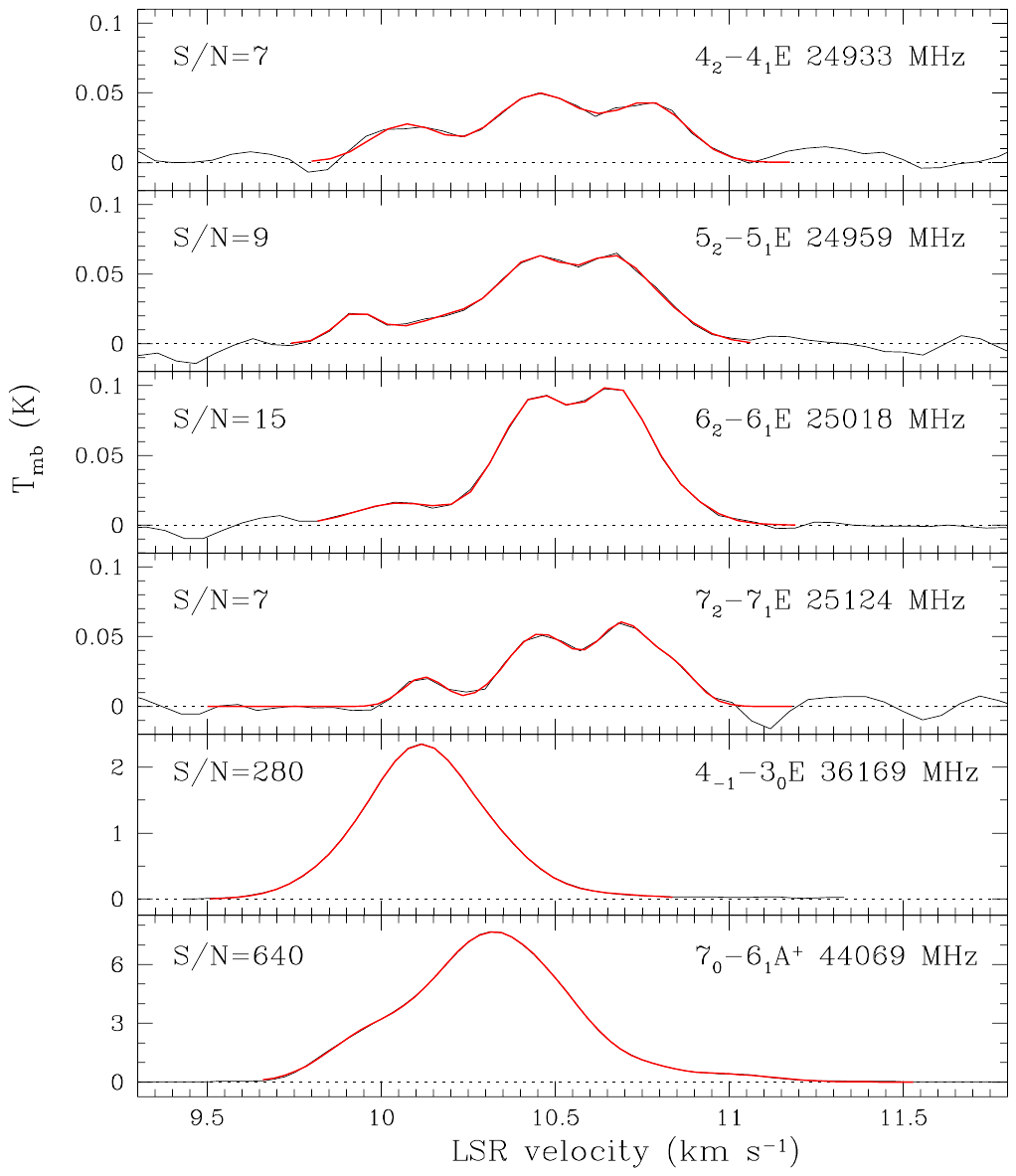}
\vspace{-2.5cm}
\caption{\small
Thermal profiles of several methanol transitions at 25~GHz observed towards RMS3865.
Strong maser lines at 36~GHz and 44~GHz are shifted relative to the thermal 25~GHz
lines by $\sim 0.5$ \kms.
The observed lines are shown in black, and the
fitting curves in red.
The central double peaks in each panel represent a partially resolved HF structure
of thermal lines (two main HF modes).
}
\label{F2}
\end{figure*}

Table~\ref{T5} shows that the splitting between two convolved HF modes of the methanol lines
at 25~GHz is $\sim 0.2$~\kms.
Being observed with a sufficiently high spectral resolution,
these modes could be distinguished in the recorded astronomical spectra,
but in practice the resolved HF modes are rare occasions due to the large Doppler broadening
which causes line merging.
For instance, among our dataset of 22 objects we detected only one system, namely, RMS3865,
with the resolved HF modes in several methanol transitions at 25~GHz.
In Fig.~\ref{F2},
the observed profiles of the $4_2-4_1E$, $5_2-5_1E$, $6_2-6_1E$, and $7_2-7_1E$
thermal emission lines are shown in black, whereas the fitting curves are depicted in red.
All lines are weak with correspondingly low S/N~$\simeq 7-15$, so a unique
deconvolution into individual subcomponents is impossible.
The parameters that can be more or less accurately estimated from the profile fitting are
solely the FWHM and the splitting $\Delta V$ between two  composite HF modes centered at
the radial velocities $V_1$ and $V_2$.
The corresponding values are given in Table~\ref{T6}.

The measured splittings $\Delta V$ coincide with those obtained in laboratory
(see Table~\ref{T5}), thus supporting our assumption
that the 25~GHz spectra towards RMS3865 represent two modes of
the composite HF transitions and are not two separate overlapping lines.
The object RMS3865 exhibits also a strong maser emission at 36 and 44~GHz with
the corresponding lines
shifted by $-0.4$~\kms\ and $-0.2$~\kms\ relative to the centre of the described 
25~GHz lines. Moreover, the kinematic models for the 36 and 44 GHz profiles are different.  
Consequently, the emissions at 36 and 44~GHz which
could correspond to the observed 25~GHz emission fall in the wings of those maser lines
and cannot be extracted, but in any case they are very weak and probably thermal.
This indicates that the source RMS3865 is in fact a superposition of several spots,
each with different composition of masering and thermally-excited species and, hence, 
with different physical conditions.
We note that interferometric observations with high spatial resolution show that such a picture
is quite common (e.g., Voronkov \etal\ 2014). 

As it was in the case of laboratory measurements, 
the observed astronomical profiles at 25~GHz towards RMS3865
show neither a decrease nor an increase in the HF mode splitting $\Delta V$ with changing $J$.
Another fact to be mentioned is that the $4_2-4_1E$ line profile itself
and parameters evaluated from its fitting support our guess in the previous section that the recorded
laboratory profile of the $4_2-4_1E$ line is indeed distorted.

\newcounter{T7}
\setcounter{T7}{7}

{\hspace{-5.0cm}
\begin{landscape}
\begin{table}
\centering
\label{T7}
\begin{tabular}{l r@.l r@.l r@.l r@.l r@.l r@.l r@.l r@.l r@.l r@.l }
\multicolumn{21}{l}{Table~\arabic{T7}.\, The line peak velocities, $V_i$, for a subsample of 
9 targets (10 maser spots) with maser activity detected in all three frequency bands at 25, 36, and}\\
\multicolumn{21}{l}{44~GHz. Listed in columns are: (1) shortened source names in accord with Table~\ref{T1}; 
(2) letters L or R indicate which component of the HF multiplet at 44~GHz  }\\
\multicolumn{21}{l}{is masering (see Fig.~\ref{F6}); L/R~-- indefinite attribution caused by a large error in the 
velocity difference $V_{44} - V_{95}$ (see text for details), no entry~-- object was }\\
\multicolumn{21}{l}{not considered previously; (3)~--(7) no entry~-- line cannot 
be extracted from the noise; 
(8) the reference velocity for the 25~GHz transitions,  $V_{\scriptscriptstyle 25}$, }\\
\multicolumn{21}{l}{calculated as a weighted mean of the peak velocities of the $5_2-5_1E$ and $6_2-6_1E$ lines, 
or as a weighed mean of other 25~GHz transitions if the position of either}\\
\multicolumn{21}{l}{$5_2-5_1E$ or $6_2-6_1E$ lines is an outlier (the second option is marked by an asterisk);
(11) the difference between the peak velocity $V_{36}$ at 36~GHz and the reference  }\\
\multicolumn{21}{l}{velocity at 25~GHz, $\Delta V_{\scriptscriptstyle 36-25} = V_{\scriptscriptstyle 36} - V_{\scriptscriptstyle 25}$;
(12) the difference between the peak velocity $V_{44}$ at 44~GHz and the reference velocity at 25 GHz,
$\Delta V_{\scriptscriptstyle 44-25} = V_{\scriptscriptstyle 44} - V_{\scriptscriptstyle 25}$. }\\
\multicolumn{21}{l}{The numbers in parenthesis are statistical errors (1$\sigma$) in the last digits.}\\
\hline\\[-10pt]
\multicolumn{1}{c}{ Source\,\, HF} &
\multicolumn{2}{c}{ $V_{\scriptscriptstyle 3_2-3_1E}$} &
\multicolumn{2}{c}{ $V_{\scriptscriptstyle 4_2-4_1E}$} &
\multicolumn{2}{c}{ $V_{\scriptscriptstyle 5_2-5_1E}$} &
\multicolumn{2}{c}{ $V_{\scriptscriptstyle 6_2-6_1E}$} &
\multicolumn{2}{c}{ $V_{\scriptscriptstyle 7_2-7_1E}$} &
\multicolumn{2}{c}{ $V_{\scriptscriptstyle 25}$} &
\multicolumn{2}{c}{ $V_{\scriptscriptstyle 36}$} &
\multicolumn{2}{c}{ $V_{\scriptscriptstyle 44}$} &
\multicolumn{2}{c}{ $\Delta V_{\scriptscriptstyle 36-25}$} &
\multicolumn{2}{c}{ $\Delta V_{\scriptscriptstyle 44-25}$} \\
 & \multicolumn{2}{c}{\footnotesize (km~s$^{-1}$)} &
\multicolumn{2}{c}{\footnotesize (km~s$^{-1}$)} &
\multicolumn{2}{c}{\footnotesize (km~s$^{-1}$)} &
\multicolumn{2}{c}{\footnotesize (km~s$^{-1}$)} &
\multicolumn{2}{c}{\footnotesize (km~s$^{-1}$)} &
\multicolumn{2}{c}{\footnotesize (km~s$^{-1}$)} &
\multicolumn{2}{c}{\footnotesize (km~s$^{-1}$)} &
\multicolumn{2}{c}{\footnotesize (km~s$^{-1}$)} &
\multicolumn{2}{c}{\footnotesize (km~s$^{-1}$)} &
\multicolumn{2}{c}{\footnotesize (km~s$^{-1}$)} \\

\multicolumn{1}{c}{\footnotesize (1)\,\,\,\,\,\,\,\,\,\,\,\,\,\,\, (2)} &
\multicolumn{2}{c}{\footnotesize (3)} &
\multicolumn{2}{c}{\footnotesize (4)} &
\multicolumn{2}{c}{\footnotesize (5)} &
\multicolumn{2}{c}{\footnotesize (6)} &
\multicolumn{2}{c}{\footnotesize (7)} &
\multicolumn{2}{c}{\footnotesize (8)} &
\multicolumn{2}{c}{\footnotesize (9)} &
\multicolumn{2}{c}{\footnotesize (10)} &
\multicolumn{2}{c}{\footnotesize (11)} &
\multicolumn{2}{c}{\footnotesize (12)} \\
\hline\\[-10pt]

{G018(1)\,\,\,  L/R} & {45}&{911(15)} & {45}&{905(6)} & {45}&{896(5)} &
{45}&{897(4)} & {45}&{906(8)} & {45}&{897(2)} & {45}&{741(6)} &
{45}&{ 882(8)} & {$-0$}&{ 156(6)} &  { $-0$}&{ 015(9)} \\

{G018(2)\,\,\, L/R} & {46}&{47(2)} & {46}&{474(10)} & {46}&{477(10)} &
{46}&{477(10)} & {46}&{460(10)} & {46}&{477(7)} & {46}&{293(20)} &
{46}&{500(18)} & {$-0$}&{184(20)} & {0}&{023(19)} \\

{B4518\,\,\,\,\,\,\,\,\,  L} & {91}&{653(7)} & {91}&{666(4)} & {91}&{652(4)} &
{91}&{639(6)} & {91}&{616(10)} & {91}&{648(4)} & {91}&{487(6)} &
{91}&{579(7)} & {$-0$}&{161(7)} & {$-0$}&{069(9)} \\

{B2147\,\,\,\,\,\,\,\,\, R} & {18}&{92(2)} & {19}&{021(2)} & {19}&{034(2)} &
{19}&{029(2)} & {19}&{015(2)} & {19}&{032(2)} & {18}&{890(4)} &
{19}&{048(2)} & {$-0$}&{142(4)} & {0}&{016(3)} \\

{B6863\,\,\,\,\,\,\,\,\, R} & {$-6$}&{56(3)} & {$-6$}&{494(6)} &
{$-6$}&{502(2)} & {$-6$}&{500(2)} & {$-6$}&{505(2)} &
{$-6$}&{501(2)} & {$-6$}&{642(10)} & {$-6$}&{488(5)} &
{$-0$}&{141(10)} & {0}&{013(5)} \\

{R3749\,\,\,\,\,\,\,\,\, R} & {8}&{815(30)} & {8}&{815(15)} & {8}&{818(11)} &
{8}&{803(11)} & \multicolumn{2}{c}{ } & {8}&{817(5)$^\ast$} & {8}&{684(8)} &
{8}&{840(4)} & {$-0$}&{133(9)} & {0}&{023(6)} \\

{B6815\,\,\,\,\,\,\,\,\, L} & {13}&{98(3)} & {14}&{010(35)} &
{14}&{048(26)} &
{14}&{023(11)} & {14}&{022(35)} & {14}&{023(11)$^\ast$}
& {13}&{928(10)} &
{14}&{090(6)} & {$-0$}&{095(15)} & {0}&{067(9)} \\

{G013 } & \multicolumn{2}{c}{ } & {43}&{633(8)} & {43}&{634(8)} &
{43}&{626(14)} & {43}&{612(18)} & {43}&{632(7)} &
{43}&{570(16)} & {43}&{749(10)} & {$-0$}&{062(18)} & {0}&{117(12)} \\

{R2879\,\,\,\,\,\,\,\,\, R} & {42}&{5(3)} & {41}&{671(14)} & {41}&{668(8)} &
{41}&{670(5)} & {41}&{663(15)} & {41}&{669(3)}  &
{41}&{44(4)} & {41}&{778(2)} & {$-0$}&{23(4)} & {0}&{109(4)} \\

{G208  } & {11}&{36(4)} & {11}&{390(13)} & {11}&{387(10)} &
{11}&{393(5)} & {11}&{384(4)} & {11}&{392(4)} &
{11}&{345(20)} & {11}&{515(18)} & {$-0$}&{047(20)} & {0}&{123(18)} \\
\hline
\end{tabular}
\end{table}
\end{landscape}
}

\refstepcounter{table}

\subsection{Sources with maser activity at 25 GHz}
\label{SSec3-5}

As stated above, the aim of the present work is to reconstruct the HF structure
of methanol transitions at 25, 36, and 44~GHz using sources with maser activity in all three families.
The presence of emission lines near 25~GHz 
with detectable maser characteristics was the main selection criterion for
the subsample considered in this section, since only for these lines we have
accurate laboratory frequencies for individual HF components
which can be used as a reference for further comparisons.
As a maser characteristic we consider here primarily the narrowness of the line
profiles. The preceding section shows that the thermally excited methanol lines
near 25~GHz have FWHM~$\ga 0.5$ \kms.
Thus, if we observe a line with FWHM~$\sim 0.25$ \kms, 
it is most likely non-thermally excited,
which could result in an inversion of the level populations.

In general, masers near 25~GHz are much less frequently observed as compared with
those at 44 and 95~GHz (Ladeyschikov \etal\ 2019).
As shown in Leurini \etal\ (2016),
the inverted level populations of transitions near 25~GHz require much
higher gas densities as compared to those at 36 and 44~GHz 
and this explains why a detection of maser activity in all three frequency bands is a relatively rare event.
We confirm this statistics by our observations as well, when among 
22 targets selected from the surveys at 44 and 95~GHz 
only 9 show the required non-thermal emission near 25~GHz.

\begin{figure*}
\vspace{-2cm}
\centering
\includegraphics[width=1.0\textwidth]{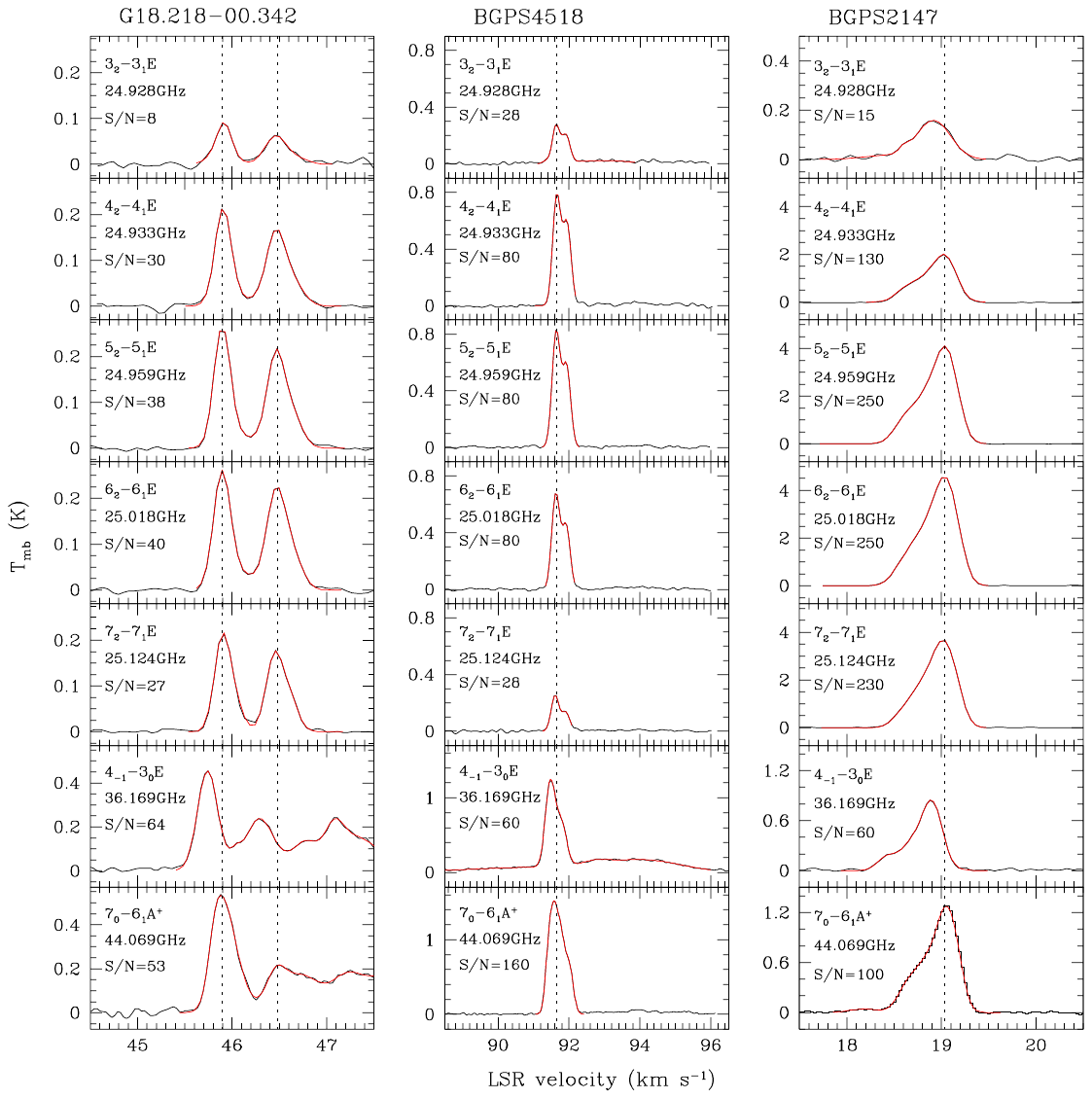}
\vspace{-3.5cm}
\caption{\small
Black curves are the baseline subtracted emission lines of Class~I CH$_3$OH masers
of three families at 25, 36, and 44~GHz observed with the Effelsberg 100-m telescope
towards the indicated maser sources.
The fitting curves are shown in red.
The methanol transition, its adopted rest frequency from Table~\ref{T2}, and
the signal-to-noise ratio (S/N) per channel
at the line peak are depicted in each panel.
The vertical dotted lines mark the average position of the $5_2-5_1E$ and $6_2-6_1E$ lines
used as the reference velocity for the 25~GHz transitions.
}
\label{F3}
\end{figure*}

\begin{figure*}
\vspace{-2cm}
\centering
\includegraphics[width=1.0\textwidth]{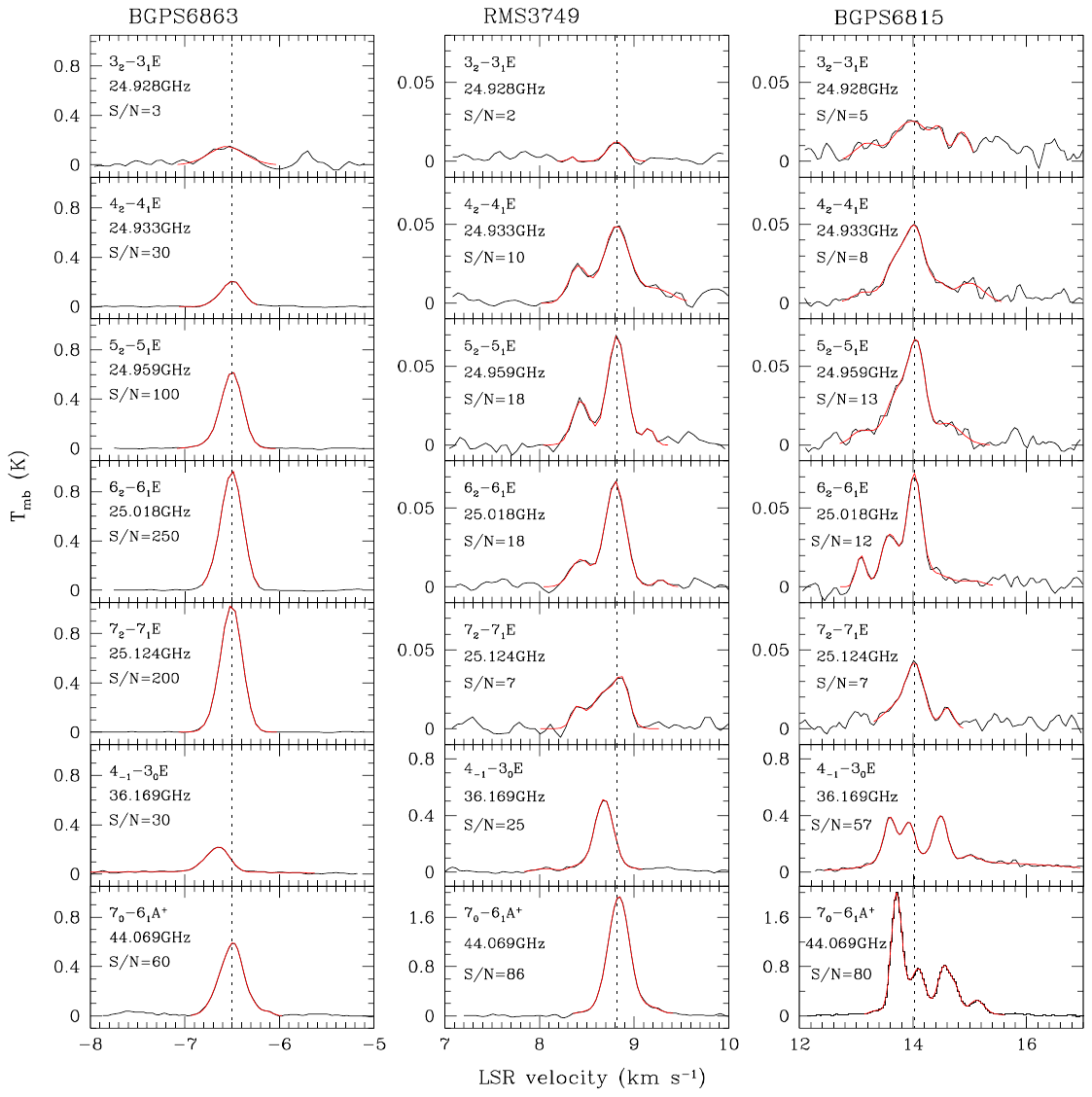}
\vspace{-3.5cm}
\caption{\small
Same as Fig.~\ref{F3}.
}
\label{F4}
\end{figure*}

\begin{figure*}
\vspace{-2cm}
\centering
\includegraphics[width=1.0\textwidth]{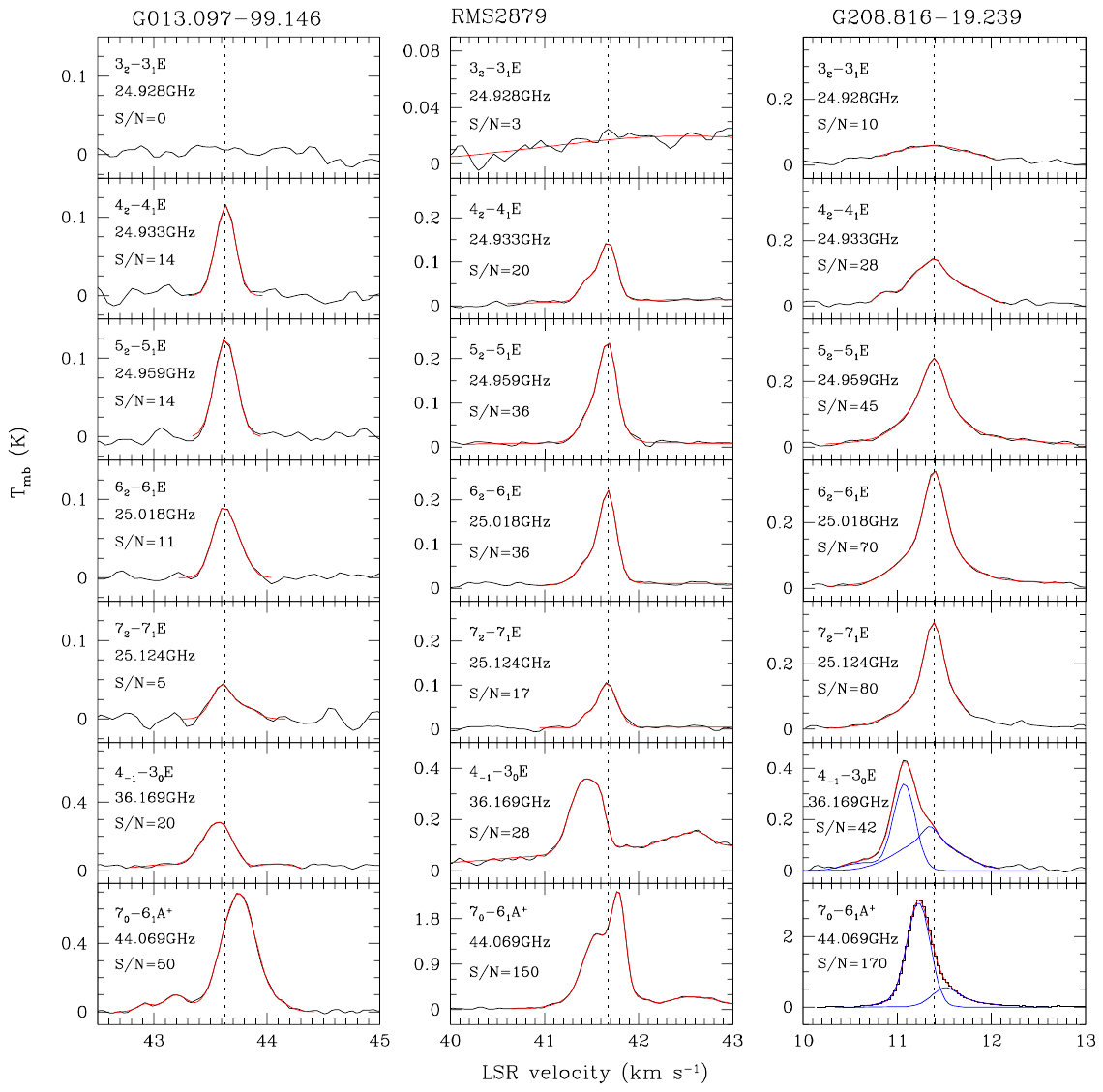}
\vspace{-3.5cm}
\caption{\small
Same as Fig.~\ref{F3}.
The blue curves in the panels $4_{-1}-3_0E$ and $7_0-6_1A^+$ in the third column show
two decomposed modes (see text).
}
\label{F5}
\end{figure*}

The emission line profiles for these 9 sources are shown in Figs.~\ref{F3}-\ref{F5}.
The line peak velocities calculated as described in Sect.~\ref{SSec3-1} are given in Table~\arabic{T7}.
Attribute R or L in the 2nd column of this table indicates which component of the HF multiplet
at 44~GHz is masering: L stands for the strongest HF component blueward
(on the velocity scale) from the torsion-rotation multiplet centre, and R, correspondingly,
for the strongest `red' HF component.
These two groups, L and R, were defined in L22 by comparing the peak
velocities of the 44 and 95~GHz maser lines (see Table~6 and Fig.~6 in L22).

Below brief comments are given on the selected 9 sources:
\begin{enumerate}
\item {\it G018.218--0.342}.
All observed transitions near 25~GHz (Fig.~\ref{F3}) show double-peaked profiles with the peaks
separated by $\sim 0.4$ \kms\
which is twice the splitting between two convolved HF modes in thermally excited lines.
The FWHM value for the first component, the same for all 25~GHz lines, is
$\sim 0.2$ \kms, and $\sim 0.28$ \kms\ for the second one.
Thus, in G018.218--0.342, emission at 25~GHz comes
from two maser spots clearly separated in the radial velocities
and with masering activity involving only half of the HF multiplet structure.

The profiles of the 36 and 44~GHz lines are much more complex and include
both the narrow maser and broad, possibly thermal components.
Indefinite attributes L/R in the 2nd column of Table~\arabic{T7} are due to
large errors in the peak velocities of the multimodal line at 95~GHz (see Fig.~5 in L22)
which makes it impossible to determine unambiguously which HF component, left or right, is masering.

\item {\it BGPS4518}. Again double-peaked maser line profiles are observed for all transitions near 25~GHz,
with a weak broad, possibly thermal contribution to the $3_2-3_1E$ line (Fig.~\ref{F3}).
A strong asymmetric maser line is present at 36~GHz, also accompanied by broad emission.
The maser line at 44~GHz shows a similar profile, but in this case without the broad component.

For all transitions :near 25~GHz, the lines have the same FWHMs of $0.55\pm0.02$ \kms\
and the same apparent splitting between the components of $0.250\pm0.010$ \kms.
This is close to the parameters measured for thermally excited lines with partly
resolved HF structure (Sect.~\ref{SSec3-3} and \ref{SSec3-4}), so 
one might have thought that here both HF modes were masering.
Yet it is not the case.
All the line profiles at 25~GHz and the maser line at 36~GHz (with removed broad emission) can be perfectly fitted
to the same two-component model, and the velocity difference between the components comes out to be equal for
all species, i.e., the 36~GHz line as a whole is shifted in frequency relative to the lines at 25~GHz.
However, the methanol HF structures at 25~GHz and 36~GHz are different
(it is more compact at 36~GHz~-- see Fig.~\ref{F6} below), 
and if several HF components at 25 and 36~GHz were
masering, then such uniform line transfer would not occur.
Thus, in BGPS4518 we observe again two separate maser spots and in every spot
merely half of the HF components (in fact only one~-- 
that with the largest Einstein A coefficient) is masering.

Unlike the 25 and 36~GHz lines, the maser at 44~GHz requires for the accurate
line profile fitting a more complex (at least 4-component) model, so that a
one-to-one correspondence between the model components at 44~GHz and those at 25 and 36~GHz cannot be established.
That is why we give in Table~\arabic{T7} only one velocity value for this object,
namely, that one which corresponds to the maximum amplitude of the line profiles.

The revealed discrepancy between the fitting models for the line profiles at 25 and 36~GHz on the one hand and for
that at 44~GHz on the other demonstrates that in the present case the
maser emission at 44~GHz does not exactly trace the maser emission at 25 and 36~GHz
probably because of the presence of an additional maser spot
radiating only at 44 and 95~GHz (see Fig.~4 in L22).
This emission may cause the velocity difference $\Delta V_{\scriptscriptstyle 44-25}$
to be significantly biased. 

\item
{\it BGPS2147}.
Among the 25~GHz transitions (Fig.~\ref{F3}), broad emission is observed in
the $3_2-3_1E$ line with FWHM~$= 0.46\pm0.04$ \kms,
whereas other lines are masers with some broad contribution as well which leads to a relatively
large value of FWHM~$= 0.400\pm 0.010$ \kms.
At 36 and 44~GHz, the maser emission, too, is accompanied by a weak broad component in the blue wings.
All maser line profiles are simple and observed with high S/N~$\ga 100$. This, in turn, results in
the high accuracy of the calculated peak velocities which are determined with errors of only
a few meters per second.
In our dataset, this object can be considered as a nearly perfect example of multi-band maser emission.

\item
{\it BGPS6863}.
Here again the $3_2-3_1E$ line is broad with FWHM~$= 0.44\pm 0.04$ \kms,
whereas all other transitions at 25~GHz are masers with very narrow line profiles
having FWHM~$= 0.260 \pm 0.005$ \kms\ (Fig.~\ref{F4}).
The maser line at 36~GHz shows extended but weak emission in the blue wing, whereas the line profile
at 44~GHz is dominated entirely by maser emission.
This object is very similar to BGPS2147 and is a second fiducial example
of the multi-band maser emission in our sample.

\item
{\it RMS3749}.
All of the present 25~GHz transitions are narrow with FWHM~$= 0.25\pm0.05$ \kms, i.e.,
we observe definitely non-thermal emission, but it is weak (Fig.~\ref{F4}).
The much stronger lines at 36 and 44~GHz, which are clearly masers, 
have the same line width. 
A weak contribution of a broad emission is also seen in the wings.

\item
{\it BGPS6815}.
At the position of the $3_2-3_1E$ line at 25~GHz, 
there is an extended emission which
is barely distinguishable from noise (Fig.~\ref{F4}).
Other 25~GHz lines represent a blend of a few narrow (FWHM~$\la 0.3$ \kms) components
which are resolved only in the strongest  $6_2-6_1E$ line.
The components are separated by $\ga 0.3$ \kms.
The profiles at 36 and 44~GHz also exhibit complex multicomponent shapes.

In the previous work (L22), we have already analyzed the 44~GHz
line together with the line at 95~GHz, but considered only the velocities of the first (strongest)
component~--
the measured difference $\Delta V_{\scriptscriptstyle 44-95} = V_{44} - V_{95}$ = 0.030 \kms\ attributed
this component as being  the masering component right to the multiplet centre (Fig.~6 in L22).
However, the detected $6_2-6_1E$ maser line at 25~GHz corresponds
to the second maximum in the line profiles at 95, 44 and 36~GHz.
Again, comparing the 44 and 95~GHz profiles (observed with the KVN),
we obtain for this second maximum a peak velocity difference of $0.008\pm0.010$ \kms\ and,
hence, classify it as being a masering component to the left of the multiplet centre.
Thus, in two neighboring components of the 44 and 95~GHz lines
we have two different HF transitions acting as masers
what confirms our assumption that in BGPS6815
a superposition of separate maser spots is observed.

\item
{\it G013.097--0.146}.
The $3_2-3_1E$ line at 25~GHz is completely buried in noise (Fig.~\ref{F5}).
Other 25~GHz transitions show weak maser emissions with
FWHM~$= 0.220\pm0.010$ \kms. A strong maser line at 36~GHz is accompanied by 
weak broad emission in both line wings.
Some broad  emission is also present in the blue wing of the maser line at 44~GHz.

\item
{\it RMS2879}.
A weak (S/N = 3) and very broad (FWHM~$ > 1$ \kms) emission is 
tentatively detected at the position of
the $3_2-3_1E$ line at 25~GHz (Fig.~\ref{F5}).
This emission is also seen in the wings of relatively strong
and narrow lines of other 25~GHz transitions which are definitely masers~--
their widths are FWHM~$= 0.25\pm0.02$ \kms.
The 36~GHz line is a combination of maser and broad, possibly thermal, emissions with a maser profile having
FWHM~$= 0.44\pm0.01$ \kms\ and showing an almost flat-topped profile.
A comparison with the maser profile at 44~GHz reveals that
the maser line at 36~GHz is in fact a blend of a few components.
The difference $\Delta V_{\scriptscriptstyle 36-25} = V_{\scriptscriptstyle 36} - V_{\scriptscriptstyle 25}$
between the peak velocities of the 36~GHz and 25~GHz lines deviates greatly from all other
$\Delta V_{\scriptscriptstyle 36-25}$ values from Table~\arabic{T7},
indicating that here we probably compare mismatched components.
However, the flat top makes the unambiguous deconvolution impossible and, hence,
the obtained $\Delta V_{\scriptscriptstyle 36-25}$ value
should be considered as an outlier.
In turn, the double-peaked maser emission line at 44~GHz is perfectly fitted to a simple
two-component model, so that the component centres can be determined with high accuracy.
To note is that here the centre of the appropriate component just corresponds
to the line peak velocity, thus making the  difference $\Delta V_{\scriptscriptstyle 44-25}$ even model-independent.

\item
{\it G208.816--19.239}.
This target resides in the Orion Molecular Cloud 2
(OMC-2) and consists of several
radiating spots overlapping in velocity space (e.g., van Terwisga \etal\ 2019).
We note that such a picture is quite common for star-forming regions and molecular
clouds  (e.g., Pagani \etal\, 2017).
The $3_2-3_1E$ line at 25~GHz is broad and probably thermal with FWHM~$ = 1.10\pm0.05$ \kms\ (Fig.~\ref{F5}).
Other 25~GHz lines represent 
a combination of the broad with the  narrow~-- maser~-- profiles with FWHM decreasing 
from 0.56 \kms\ for the $4_2-4_1E$ line to 0.30 \kms\ for the $7_2-7_1E$ line.
In the profiles of the 36 and 44~GHz lines, a strong maser emission from other spot(s) 
is present, which manifests in large
differences between the peak velocities of the transitions near 25~GHz and those at
36 and 44~GHz: $\sim 0.3$ and $\sim 0.45$ \kms, respectively.
However, unlike RMS3865 (Section~\ref{SSec3-4}), the latter profiles show a pronounced 
asymmetry in the red wings
and can be decomposed into two (multicomponent) modes using {\it the same}
fitting model. This fact together with a high S/N makes it possible to calculate the centres of both modes with
an accuracy sufficient to utilize the corresponding differences $\Delta V_{\scriptscriptstyle 36-25}$ and
$\Delta V_{\scriptscriptstyle 44-25}$ in the subsequent analysis.
The synthetic profiles of the two extracted modes are shown in blue in Fig.~\ref{F5}.
\end{enumerate}

\section{HF structure of CH$_3$OH as recovered from 25, 36, and 44~GHz maser spectra}
\label{Sec4}

\subsection{`Favored' HF components in the 25, 36, and 44~GHz torsion-rotation multiplets}
\label{SSec4-1}

The procedure described in this section is based on the assumption that in any one maser source
only one HF component of the corresponding torsion-rotation
multiplet is masering and this component can change from source to source. 
In order to determine which component is masering
we employ the multiplet HF structure obtained either from laboratory 
measurements (for masers at 25~GHz) or
from quantum-chemical calculations (for masers at 36 and 44~GHz) and consider the differences between the
peak velocities of the observed 25, 36 and 44~GHz maser lines, $\Delta V_{\scriptscriptstyle 36-25} = V_{36} - V_{25}$ and
$\Delta V_{\scriptscriptstyle 44-25} = V_{44} - V_{25}$. Since there are only a few (from 2 to 4) `favored' (i.e., which
can act as masers; see below) HF transitions in every HF multiplet, the values $\Delta V_{\scriptscriptstyle 36-25}$ and
$\Delta V_{\scriptscriptstyle 44-25}$ are expected to form separate groups which are then analyzed to
identify the HF components in action.

The velocity differences $\Delta V_{\scriptscriptstyle 36-25}$ and $\Delta V_{\scriptscriptstyle 44-25}$ are given in
Col.~11 and Col.~12 of Table~\arabic{T7}. For the reference velocity of 
the 25~GHz masers (Col.~8 of Table~\arabic{T7}, $V_{25}$)
the weighted mean of the peak velocities of the $5_2-5_1E$ and $6_2-6_1E$ lines is taken, since
just these two lines are the strongest in our dataset, thus allowing us to calculate the line
centres with the highest possible accuracy. On the other hand, the measured mean velocity difference between the
$5_2-5_1E$ and $6_2-6_1E$ lines of only $0.002\pm0.002$ \kms\ indicates that their positions coincide.

\begin{figure*}
\vspace{-5cm}
\centering
\includegraphics[width=1.0\textwidth]{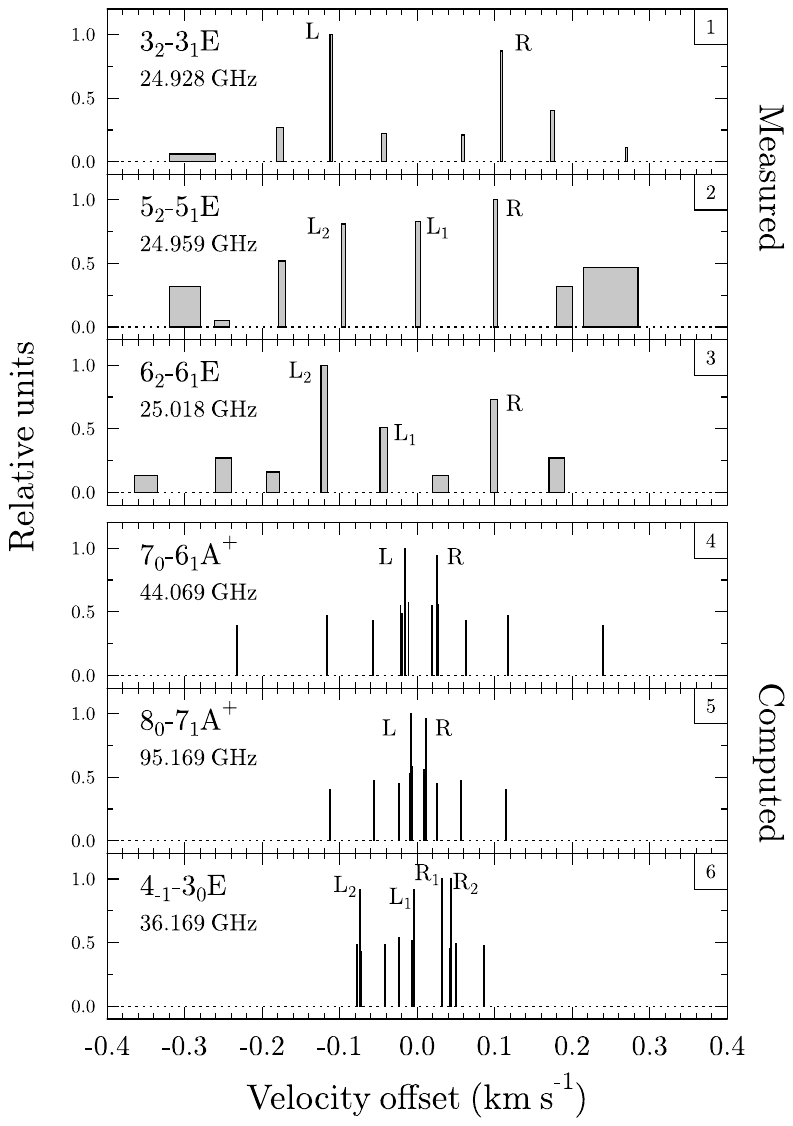}
\vspace{-0.5cm}
\caption{\small
Templates of the HF structure of torsion-rotation multiplets
used in the present study.
Individual HF components are represented by bars with bar's height
proportional to the observed amplitude (the upper three panels) or to
the Einstein A-coefficient (the lower three panels) of the corresponding transition.
The velocity zero point is the multiplet centre.
Indicated with letters are the `favored' HF components which potentially
can form a maser.
{\it Panels 1-3:} HF components deconvolved from the laboratory spectra (Fig.~\ref{F1}
and Table~\ref{T4}); the bar widths indicate the  $\pm1\sigma$ uncertainty interval of the line position.
{\it Panels 4-6:} results of quantum-chemical calculations of the methanol HF structure (Lankhaar \etal\ 2016).
}
\label{F6}
\end{figure*}

Figure~\ref{F6} schematically shows the HF components of different torsion-rotation transitions
in methanol considered in this section. 
The zero velocity corresponds to the central frequency of the specified multiplet as given in Table~\ref{T2}.

The hyperfine component structure of the $3_2-3_1E$, $4_2-4_1E$, $5_2-5_1E$, and $6_2-6_1E$ transitions
near 25~GHz was calculated from the laboratory spectral profiles 
(see Section~\ref{SSec3-3}).
The individual HF components are plotted as bars at the velocity offsets $\Delta V$ given in Table~\ref{T4} with
bar's  height scaling as the amplitude of the corresponding component,
and bar's width~-- as the error in $\Delta V$. 
The amplitude (emissivity), which is proportional to the Einstein A coefficient,
determines the optical depth $\tau$ of the corresponding transition.
Taking into account that maser emission exponentially depends on $\tau$,
it becomes clear why the HF components with the largest Einstein A coefficients 
are considered as `favored' for maser action. 

As noted in
Section~\ref{SSec3-3}, some bars in fact represent several convolved components; this is particularly true
for the prominent bar to the 
right of the multiplet centre (indicated with R in Fig.~\ref{F6}) which is probably a close doublet
with the splitting between components less than 0.010 \kms. 
On the other hand, strong components to the left of the centre
form a broad doublet (best seen in the $5_2-5_1E$ transition with components indicated as L$_1$ and L$_2$) with
a splitting of $\sim 0.1$ \kms\  and with one component very close to the multiplet centre.

The available laboratory measurements for
the torsion-rotation transitions $7_0-6_1A^+$  at 44~GHz and $8_0-7_1A^+$ at 95~GHz are not fine enough
to resolve the HF components (Tsunekawa \etal\ 1995; M\"uller \etal\ 2004), that is why we use here
the HF structures computed by the quantum-chemical model of Lankhaar \etal\ (2016, 2018).
Plotted by bars are only the strongest HF components, i.e., those with large Einstein A coefficients,
with bar's height proportional to the corresponding coefficient.
In general, computed structures are strongly model-dependent and require in every case experimental verification.
As already noted in Section~\ref{Sec1}, by analyzing previous observations of the
44 and 95~GHz masers in $A$-methanol (L22), we found that computed velocity
offsets, $\Delta V$, for the strongest (i.e., masering) HF components to the right (R)
and to the left (L) of the multiplet centres correctly reproduce the observational data, i.e., for these
transitions the model of Lankhaar \etal\ is
adequate and can be utilized as a template.
At 44~GHz, the computed offsets (on the velocity scale) are
$\Delta V_{\scriptscriptstyle\rm R} = 0.026$ \kms\ and
$\Delta V_{\scriptscriptstyle\rm L} = -0.016$ \kms, thus giving the splitting between the R and L components
$\Delta V^{\scriptscriptstyle\rm R-L}_{\scriptscriptstyle 44} = 0.042$ \kms.
However, it is to emphasize once again that offsets (in frequencies or velocities) are calculated
{\it relative} to the multiplet centre which is simply an external model parameter.
In order to obtain the absolute (i.e., zero-point independent) frequencies of individual HF components
an accurate value of the central rest frequency is needed.

The HF structure of the $4_{-1}-3_0E$ torsion-rotation multiplet at 36~GHz (panel 6 in Fig.~\ref{F6})
was also computed by the quantum-chemical model mentioned above.
At the moment we have neither laboratory nor astronomical data which can be used to
verify the calculations. We note only that  the computed structure looks very
similar to that reconstructed from the laboratory spectra of the $E$-methanol transitions at 25~GHz with four
potentially `favored'(strong) components: two close components (not
resolved explicitly at 25~GHz) to the right of the multiplet centre (indicated as R$_1$ and R$_2$
in panel 6 of Fig.~\ref{F6})
and two distant components to the left (L$_1$ and L$_2$ in panel 6 of Fig.~\ref{F6}).
This structure differs significantly from the HF structure at 44 and 95~GHz in $A$-methanol
where only two favored components~-- one to the right and one to the left of
the centre~-- are present (and confirmed by astronomical observations).

\begin{figure*}
\vspace{-2cm}
\centering
\includegraphics[width=1.0\textwidth]{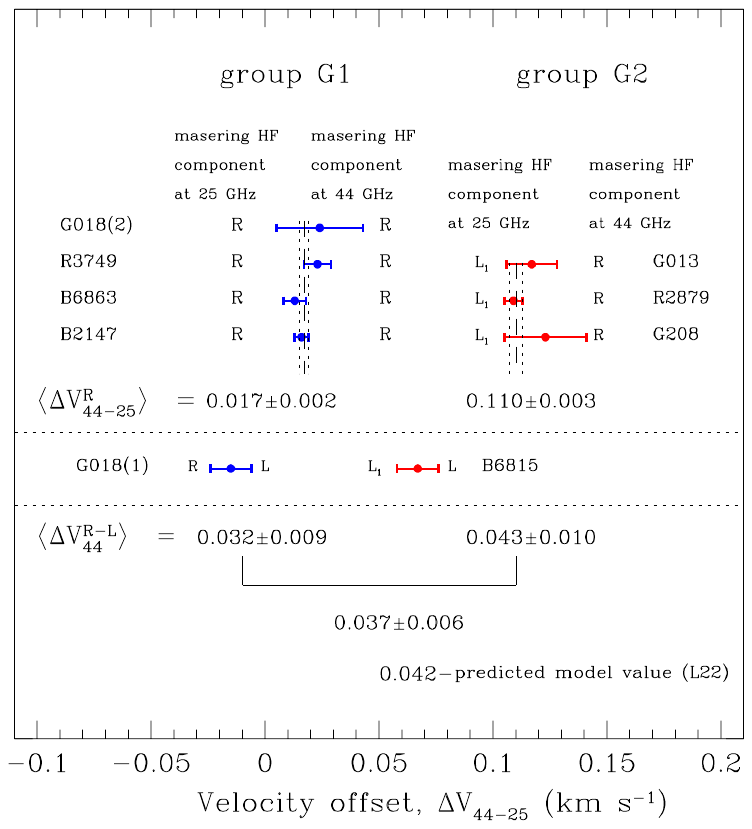}
\vspace{-6cm}
\caption{\small
The difference $\Delta V_{\scriptscriptstyle 44-25} = V_{44} - V_{25}$ between the
peak velocities of maser lines
at 25 and 44~GHz. The masering HF components labeled by
L$_1$ and R at 25~GHz and L and R at 44~GHz are as indicated in  
panels~2, 3 and 4 in Fig.~\ref{F6}. See text for details.
}
\label{F7}
\end{figure*}

The measured velocity differences,
$\Delta V_{\scriptscriptstyle 44-25} = V_{44} - V_{25}$, are
plotted as points with error bars in Fig.~\ref{F7}.
It is seen that the
$\Delta V_{\scriptscriptstyle 44-25}$ points with label R at 44~GHz,
i.e., those with maser emission due to the {\it same} right-handed HF component,
form, nevertheless, {\it two} compact and clearly distinguishable groups (referred to as G1 and G2):
the G1 group with a weighted mean
$\langle\Delta V^{\scriptscriptstyle\rm R}_{\scriptscriptstyle 44-25}\rangle_{\scriptscriptstyle\rm G1}
= 0.017\pm0.002$ \kms,
and the G2 group with
$\langle\Delta V^{\scriptscriptstyle\rm R}_{\scriptscriptstyle 44-25}\rangle{\scriptscriptstyle\rm G2}
= 0.110\pm0.003$ \kms.
These mean values together with their $\pm1\sigma$ boundaries are
indicated in Fig.~\ref{F7} with, respectively, dashed and dotted lines.
The targets G013 and G208, shown in this figure, were not included in our previous dataset and, accordingly,
it was not known in advance whether they should have been marked with letters R or L.
However, due to the closeness of their $\Delta V_{\scriptscriptstyle 44-25}$ values to that of R2879
from the G2 group, we attribute G013 and G208 to the group G2 as well (i.e., 
the masering component at 44~GHz is R).

These two groups, G1 and G2, imply that 25~GHz masers
are also formed by the emission at {\it two} different frequencies within the 25~GHz multiplets.
The offsets of these frequencies relative to the multiplet centre can be assessed as follows.
The separation between groups is
\begin{equation}
\langle\Delta V^{\scriptscriptstyle\rm R}_{\scriptscriptstyle 44-25}\rangle{\scriptscriptstyle\rm G2} -
\langle\Delta V^{\scriptscriptstyle\rm R}_{\scriptscriptstyle 44-25}\rangle_{\scriptscriptstyle\rm G1} =
0.093\pm0.004\,\,\, {\rm km~s}^{-1}\, ,
\label{Eq4-1}
\end{equation}
and, taking into account that the observed maser emission at 44~GHz is due to the same R-component of the HF multiplet,
this value should correspond to the velocity splitting
between two HF components of the 25~GHz transition.
Looking at panels 3 and 4 of Fig.~\ref{F6} where the HF structure of these multiplets is plotted
one finds that the measured splitting of $\sim 0.093$ \kms\ can be explained
if we assume that the emerging groups are formed
either ($i$) by the L$_2$ and L$_1$ components (splitting $0.096\pm0.008$ \kms) or
($ii$) by the R and L$_1$ components of the multiplets (splitting $0.100\pm0.008$ \kms). 
Note that the splitting
between the extreme components L$_2$ and R is approximately 0.2 \kms, i.e.,
twice the value obtained in Eq.~\ref{Eq4-1}.
However, in the G1 group (Fig.~\ref{F7}) we have two targets,
B2147 (Fig.~\ref{F3}) and B6863 (Fig.~\ref{F4}),
both with broad and weak $3_2-3_1E$ lines and strong narrow maser lines in the other 25~GHz transitions.
For these targets, the velocity difference between the centres of the maser and broad lines is
$0.05\pm0.03$ \kms\ and $0.10\pm0.03$ \kms, respectively,
i.e., the maser emission is formed definitely redward to the centre of the broad profile.
This means that the second option is realized, namely, that 
in the G1 group the masering component of the 25~GHz maser is R
and in the G2 group~-- L$_1$  (see diagram in Fig.~\ref{F7}).

We would like to also emphasize
that in these L$_1$ masers just a single L$_1$ component radiates
and not components L$_2$ and R together as one might think.
These components are almost equally detached
from the multiplet centre (Fig.~\ref{F6}) and could therefore form a line centered at L$_1$.
However, this line would have an FWHM~$\ga  0.5$ \kms\ (see Table~\ref{T6}),
whereas in the L$_1$ masers the observed lines are very narrow, FWHM~$\sim 0.2-0.3$ \kms,
which can be realized only for a single masering component.

In our dataset, there are only three targets, G018(1), B4518, and B6815,
with the left-handed (L) masering component at 44~GHz (Table~\arabic{T7}), and
we have to determine which of the possible HF components are masering at 25~GHz.

The values $\Delta V_{\scriptscriptstyle 44-25}$ for these sources
are $-0.015\pm0.009$ \kms\ (G018(1)), 
$-0.069\pm0.009$ \kms\ (B4518), and $0.067\pm0.009$ \kms\ (B6815).
According to the quantum-chemical model by Lankhaar \etal\ (2016, 2018), the velocity splitting
between the `favored' HF components at 44~GHz should be
$\Delta V^{\scriptscriptstyle\rm R-L}_{\scriptscriptstyle 44} = 0.042$ \kms\
(Fig.~\ref{F6}).
Comparing these values with $\Delta V_{\scriptscriptstyle 44-25}$ for the groups G1 and G2, it is
evident that in G018(1) the masering component at 25~GHz is R and in B6815 is L$_1$.
As for B4518, the masering component at 25~GHz is also R, but the value of 
$\Delta V_{\scriptscriptstyle 44-25}$ is  highly distorted due to some 
intervening emission at 44~GHz
as it was already suggested in the description of this source in Sect.~\ref{SSec3-5}. 
Consequently, we exclude $\Delta V_{\scriptscriptstyle 44-25}= -0.069$ \kms\ from 
further considerations. 
Then the velocity splitting between the R and L hyperfine components in the 44~GHz multiplet is
$\Delta V^{\scriptscriptstyle\rm R-L}_{\scriptscriptstyle 44} = 0.032\pm0.009$ \kms\
and
$\Delta V^{\scriptscriptstyle\rm R-L}_{\scriptscriptstyle 44} =
0.043\pm0.010$ \kms\ for the groups G1 and G2, respectively, yielding the combined weighted mean
$\langle \Delta V^{\scriptscriptstyle\rm R-L}_{\scriptscriptstyle 44} \rangle = 0.037\pm0.006$ \kms.
This estimate is quite close (the difference lies within the $\pm1\sigma$ uncertainty interval) to
the predicted value of 0.042 \kms\ which can be considered as a good coincidence, especially
taking into account that for each group only one target  with the left-handed
masering component at 44~GHz is available. 

\begin{figure*}
\vspace{-2cm}
\centering
\includegraphics[width=1.0\textwidth]{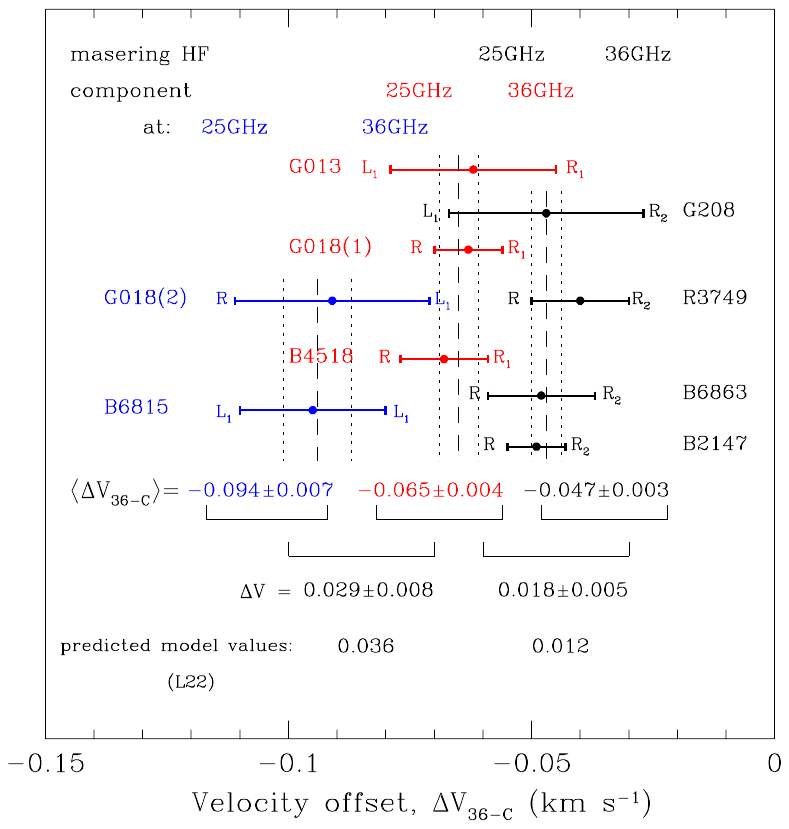}
\vspace{-6cm}
\caption{\small
The peak velocities of the 36~GHz line relative to the common reference
velocity $V_{\scriptscriptstyle C}$ of the 25~GHz lines which accounts for the
masering HF component at 25~GHz as they are shown in Fig.~\ref{F7}.
Groups with matching masering HF components of the
36~GHz multiplet are marked in different colors.
See text for details.
}
\label{F8}
\end{figure*}

Now consider the velocity difference between the 36~GHz and 25~GHz lines, $\Delta V_{\scriptscriptstyle 36-25}$.
From the above analysis of $\Delta V_{\scriptscriptstyle 44-25}$ we know for each of our targets
which HF component is masering at 25~GHz, so that the difference $\Delta V_{\scriptscriptstyle 36-25}$ can be
led to a common reference point. With the L$_1$ component in the 25~GHz multiplet
(coinciding with the centre of the multiplet, see panel 2 in Fig.~\ref{F6})
taken as such a reference, the re-calculated values $\Delta V_{\scriptscriptstyle 36-25}$ are plotted as points
with error bars in  Fig.~\ref{F8}. This time three separate groups are formed, each
characterized by the values (weighted means) of
$-0.094\pm0.007$ \kms,  $-0.065\pm0.004$ \kms, and  $-0.047\pm0.003$ \kms. These values are given relative
to the {\it same reference}, indicating that every group in fact corresponds to a certain masering component at 36~GHz with
differences between the groups indicating the separations between components.
The splitting between the first and second components is then $0.029\pm0.008$ \kms\
and between the second and third is  $0.018\pm0.005$ \kms\ (Fig.~\ref{F8}).
Comparing these estimations with the computed HF structure at 36~GHz (panel 6 in Fig.~\ref{F6}) we
easily identify the first group with the component L$_1$,
the second~-- with R$_1$ (the predicted splitting R$_1$--L$_1$ is 0.036 \kms)
and the third~-- with R$_2$
(the predicted splitting R$_2$--R$_1$ is 0.012 \kms). 
Note that the computed splitting between the components R$_2$ and R$_1$ of 0.012 \kms\ is
underpredicted~-- such small difference simply would not be
detected with our resolution and measurement errors~-- but
in general we can conclude that the quantum-chemical model by Lankhaar \etal\ (2016, 2018)
quite correctly reproduces the structure of the `favored' HF components
in the $4_{-1}-3_0E$ torsion-rotation multiplet at 36~GHz of $E$-methanol.

\subsection{Comments on the recovered HF components in the 25, 36, and 44~GHz multiplets}
\label{SSec4-2}

As already noted above, our basic assumption is that each specific multiplet has only
a few `favored' HF components which can participate in maser action and only one of them 
acts in any given source.
Using only kinematic characteristics of the maser line profiles (namely, peak velocities) and
employing templates of the multiplet's HF structure we determined which particular HF component
is acting in every case. Here, we comment on possible physical mechanisms behind the revealed effects.

Hyperfine splitting structure in $E$-methanol
at 25~GHz was recorded in laboratory experiments 
and then confirmed by observations of the thermally excited emission lines towards RMS3865.
According to this structure, components with almost equally strong Einstein A coefficients are
L$_2$, L$_1$, and R (panel 3 in Fig.~\ref{F6}).
In our dataset of 10 masers (9 targets), we detected 6 cases where
the R component was masering, 4 cases with the L$_1$ component and none with the masering component L$_2$.
A similar situation exists with the component L$_2$  in the $4_{-1}-3_0E$ multiplet at 36~GHz
of $E$-methanol (panel 7 in Fig.~\ref{F6}):
there are four `favored' components (L$_1$, L$_2$, R$_1$, R$_2$), but only three of them
(L$_1$, R$_1$, R$_2$) are observed as masers.
It appears that the probability to form a maser is the lowest just for the high-frequency component
among the `favored' ones.

In general, the probability of the stimulated emission, $P(i\to j)$, is proportional to 
the Einstein coefficient B~$\sim$~A$/f^3$, where $f$ is the transition frequency. 
Then, for a given HF transition 
$P(i\to j) \sim (A_{ij}/f^3_{\scriptscriptstyle 0})(1 - 3\Delta f/f_{\scriptscriptstyle 0})$,
and $\Delta f$ is the frequency shift between the HF transition and the multiplet centre 
$f_{\scriptscriptstyle 0}$,
i.e., with increasing $\Delta f$ the probability decreases.
However, in our case $\Delta f \sim$ 10~kHz and $f{\scriptscriptstyle 0} \sim$ 10s~GHz which 
makes the ratio $\Delta f/f_{\scriptscriptstyle 0}$ extremely small. 
Thus, to explain the revealed absence of high-frequency masering components 
in the HF multiplets by the Einstein B coefficients requires a very fine tuning of the 
masering processes which is hard to imagine. More likely is that some pumping peculiarity is responsible for this. 

In our previous study, by analyzing the 44 and 95~GHz masers in $A$-methanol, we found
that the masering components at these frequencies
were locked, i.e., the components occurred in combinations, either R-R or L-L (L22).
Now dealing with the 25 and 36~GHz masers in $E$-methanol, we do not see any component locked~--
all possible combinations of masering
components are realized (Fig.~\ref{F8}), except high-energy components 
(L) discussed above.

In order to be formed, a maser requires firstly the
inverted population of specific levels and then special physical conditions
in the medium where the beam propagates.
Whether a single masering HF component is selected by 
some kind of anisotropic pumping, or by some characteristics of
the gain (ambient) medium, or by both factors remains at the moment completely unclear.
More investigations are needed to reveal all the
physical processes behind the cosmic masers.

\begin{table}
\centering
\caption{Velocity corrections $\Delta V^{\rm cor}$ for methanol transitions
at 44 and 36 GHz calculated by Eq.(\ref{Eq5-5}).
The $1\sigma$ uncertainties in the last digits are given in parentheses.
}
\label{T8}
\begin{tabular}{l r@.l r@.l }
\hline\\[-10pt]
\multicolumn{1}{c}{Source} & \multicolumn{2}{c}{$\Delta V^{\rm cor}_{\scriptscriptstyle 44}$}  &
\multicolumn{2}{c}{$\Delta V^{\rm cor}_{\scriptscriptstyle 36}$}  \\
 & \multicolumn{2}{c}{(km~s$^{-1}$)} & \multicolumn{2}{c}{(km~s$^{-1}$)} \\
\hline\\[-10pt]
B2147 & $-0$&083(3) &  0&092(6)\\
B6863 & $-0$&080(5) &  0&091(11)\\
R3749 & $-0$&090(6) & 0&083(10)\\
G018(2) & $-0$&090(19) &   0&087(20)\\
G018(1) & $-0$&094(9) & 0&088(9)\\
G013    & $-0$&091(12) &    0&087(19)\\
R2879  & $-0$&083(4) & \multicolumn{2}{c}{ } \\
G208   & $-0$&097(18) &   0&090(20)\\
B6815  & $-0$&083(9) & 0&091(15)\\
B4518  & \multicolumn{2}{c}{ }  &    0&093(9)\\[2pt]
{\footnotesize\it weighted mean:} & $-0$&0842(13) & 0&0899(11)  \\
\hline\\[-8pt]
\end{tabular}
\end{table}

\section{Correction of the rest frequencies for the $4_{-1}-3_0E$, $7_0-6_1A^+$, and $8_0-7_1A^+$
methanol transitions}
\label{Sec5}

In the present study and in L22 we analyzed Class~I masers in
the $7_0-6_1A^+$ at 44~GHz and $8_0-7_1A^+$ at 95~GHz transitions of $A$-methanol
together with the $4_{-1}-3_0E$ transition at 36~GHz and several transitions at 25~GHz of $E$-methanol
observed towards the same Galactic targets.
The rest frequencies (centres of multiplets) of the transitions
at 25~GHz are measured in the laboratory with an accuracy higher than 1 kHz, 
whereas all other rest frequencies are known
with  larger uncertainties of $\sim 10$~kHz (Table~\ref{T2}).
Class~I methanol masers arise in compact (linear scale $\la 100$~AU) regions of a cold and quiet gas
with small velocity gradients. Maser line profiles are
simple, narrow and stable. It is natural to suppose that all observed masers originate in the same velocity
field and, hence, all line positions should coincide. Then the poorly known rest frequencies can be corrected
via the common procedure, where lines with frequencies to be adjusted are aligned with some reference line
(e.g., Dore \etal\ 2004; Pagani \etal\ 2009; Voronkov \etal\ 2014).

As a reference we chose the $5_2-5_1E$ line at 25~GHz for reasons already mentioned above: this line is
strong, so its position (peak velocity) can be calculated with small errors, and its laboratory profile
with a partly resolved HF structure had spacing and S/N ratio high enough
to estimate the frequencies of individual HF components more or less accurately.

From the analysis performed in the previous Section we know which HF component of the 25~GHz 
is masering in each source.
According to the calculations carried out with the laboratory profiles, the component L$_1$
coincides within the measurement errors (0.006 \kms) with the multiplet centre and
the component R is detached from it by $0.100\pm0.006$~\kms, so that
the difference between these components is $0.100\pm0.008$~\kms\ (Table~\ref{T4}).
From the observed maser profiles we measure $0.093\pm0.004$~\kms.
This is a more accurate estimate, especially taking into account that in the laboratory
profile the component R represents probably a close blend of two or more sub-components, so that the position of
a sub-component which is in fact masering can be slightly shifted from the position of the blend centre.
Thus, in the following calculations we
consider the HF component L$_1$ as coinciding with the $5_2-5_1E$ multiplet centre and the HF component R~--
detached from it by $0.093\pm0.004$~\kms.

For the $7_0-6_1A^+$ transition at 44~GHz and $8_0-7_1A^+$ transition at 95~GHz we take the shifts of the
masering components R and L relative to the corresponding multiplet centres as computed by 
the quantum-chemical model of Lankhaar \etal\ (2016, 2018).
The correctness of these shifts was confirmed observationally both in L22 and in the present study.

The situation with the $4_{-1}-3_0E$ multiplet at 36~GHz is more ambiguous. We measure shifts of three
masering components relative to each other and these shifts coincide within errors with those calculated
by the model. However, we have no data to estimate whether the computed shifts of the HF components
relative to the multiplet centre are also correct or not. Taking into account that in general the model
described the HF structure of the $4_{-1}-3_0E$ multiplet quite properly, we attribute for the component
L$_1$ the shift relative to the multiplet
centre as computed by the model and adjust the shifts of other components R$_1$ and R$_2$ accordingly.

The frequency correction based on the line alignment is conveniently performed on the velocity scale.
Assuming the coinciding positions of all our maser lines, we can write the following equation:
\begin{equation}
V_{\scriptscriptstyle 25} + \Delta V^{\scriptscriptstyle\rm HF}_{\scriptscriptstyle 25} =
V_{\scriptscriptstyle X} + \Delta V^{\scriptscriptstyle\rm HF}_{\scriptscriptstyle X} +
\Delta V^{\rm cor}_{\scriptscriptstyle X}\, .
\label{Eq5-2}
\end{equation}
Here the reference velocity, $V_{\scriptscriptstyle 25}$, and the velocity of the corrected species
X, $V_{\scriptscriptstyle X}$,
are measured from the observed sky frequencies using radio convention and are given in Table~\arabic{T7}.
The values of $\Delta V^{\scriptscriptstyle\rm HF}_{\scriptscriptstyle 25}$ and
$\Delta V^{\scriptscriptstyle\rm HF}_{\scriptscriptstyle X}$ are the shifts relative
to the multiplet centre of the masering HF components
(indicated for our targets in Figs.~\ref{F7} and \ref{F8}),
and $\Delta V^{\rm cor}_{\scriptscriptstyle X}$ is the velocity correction for
the adjusted X line.

The corresponding frequency correction,
$\Delta f^{\rm cor}_{\scriptscriptstyle X}$ is then calculated as
\begin{equation}
\Delta f^{\rm cor}_{\scriptscriptstyle X} =
\frac{f_{\scriptscriptstyle X}\cdot{\Delta V^{\rm cor}_{\scriptscriptstyle X}}/{c}}{1 -
{V_{\scriptscriptstyle X}}/{c} - {\Delta V^{\rm cor}_{\scriptscriptstyle X}}/{c}} \approx
f_{\scriptscriptstyle X}\cdot{\Delta V^{\rm cor}_{\scriptscriptstyle X}}/{c}\, ,
\label{Eq5-5}
\end{equation}
taking into account that the quantities $V_{\scriptscriptstyle X}/c \ll 1$, and
$\Delta V^{\rm cor}_{\scriptscriptstyle X}/c \ll 1$.

The calculated velocity corrections, $\Delta V^{\rm cor}_{\scriptscriptstyle X}$, 
for the methanol transitions at 44 and 36~GHz are listed in Table~\ref{T8}. 
The low dispersion of the obtained values in both cases indicates that the masering 
HF components and their shifts were estimated properly.

The resulting means $\Delta V^{\rm cor}_{\scriptscriptstyle 44} = -0.0842\pm0.0013$~\kms\
and $\Delta V^{\rm cor}_{\scriptscriptstyle 36} = 0.0899\pm0.0015$~\kms\ give
the frequency corrections
$\Delta f^{\rm cor}_{\scriptscriptstyle 44} = -0.0124\pm0.0002$~MHz and
$\Delta f^{\rm cor}_{\scriptscriptstyle 36} = 0.0108\pm0.0002$~MHz.

In L22, we calculated the velocity difference $\Delta V_{\scriptscriptstyle 44-95}$ for 19 Galactic targets.
Using the adjusted rest frequency for the transition at 44~GHz, we can now correct the rest frequency of the
$8_0-7_1A^+$ transition at 95~GHz.
The average of the velocity corrections given in Table~\ref{T9} is
$\Delta V^{\rm cor}_{\scriptscriptstyle 95} = -0.0679\pm0.0009$~\kms\ with
corresponding correction in frequency
$\Delta f^{\rm cor}_{\scriptscriptstyle 95} = -0.0216\pm0.0003$~MHz.

The frequency of our reference line $5_2-5_1E$ is known
with an error of 0.4~kHz (Table~\ref{T2}) which can be considered as a systematic error.
Taking this into account,
we obtain the following corrected frequencies:
$f_{\scriptscriptstyle 36} = (36169.2488\pm 0.0002_{\scriptscriptstyle\rm stat} \pm
0.0004_{\scriptscriptstyle\rm sys})$~MHz,
$f_{\scriptscriptstyle 44} = (44069.4176\pm0.0002_{\scriptscriptstyle\rm stat} \pm    
0.0004_{\scriptscriptstyle\rm sys})$~MHz,
and
$f_{\scriptscriptstyle 95} = (95169.4414\pm 0.0003_{\scriptscriptstyle\rm stat} \pm
0.0004_{\scriptscriptstyle\rm sys})$~MHz. 

\begin{table}
\centering
\caption{Velocity correction for the difference
$\left( V^{\scriptscriptstyle\rm P}_{\scriptscriptstyle 44} -
V^{\scriptscriptstyle\rm M}_{\scriptscriptstyle 95}\right)$
measured in L22
for methanol transitions at 44 and 95 GHz.
The $1\sigma$ uncertainties in the last digits are given in parentheses.
}
\label{T9}
\begin{tabular}{l r@.l c r@.l }
\hline\\[-10pt]
\multicolumn{1}{c}{Source} &
\multicolumn{2}{c}{$\Delta V^{\scriptscriptstyle\rm P}_{\scriptscriptstyle 44} -
V^{\scriptscriptstyle\rm M}_{\scriptscriptstyle 95}$}  & &
\multicolumn{2}{c}{$\Delta V^{\rm cor}_{\scriptscriptstyle 44} -
V^{\scriptscriptstyle\rm M}_{\scriptscriptstyle 95}$}  \\
 & \multicolumn{2}{c}{(km~s$^{-1}$)} & & \multicolumn{2}{c}{(km~s$^{-1}$)} \\
\hline\\[-10pt]
R149 & $0$&005(12) & L &  $-0$&072(12)\\
R153 & $0$&024(25) & R &  $-0$&08(3)\\
R2837 & $0$&031(40) & R &  $-0$&07(4)\\
R2879 & $0$&036(5) & R &  $-0$&064(5)\\
R3659 & $0$&013(15) & L &  $-0$&064(15)\\
R3749 & $0$&029(8) & R &  $-0$&071(8)\\
R3865 & $0$&033(8) & R &  $-0$&067(8)\\
B4252 & $0$&008(4) & L &  $-0$&069(4)\\
B1584 & $0$&008(30) & L &  $-0$&07(3)\\
B7501 & $0$&035(11) & R &  $-0$&065(11)\\
B7022 & $0$&025(8) & R &  $-0$&075(8)\\
B2147 & $0$&046(20) & R &  $-0$&05(2)\\
B4518 & $0$&014(18) & L &  $-0$&063(18)\\
B6518 & $0$&030(8) & R &  $-0$&070(8)\\
B6820(1) & $0$&032(19) & R &  $-0$&068(19)\\
B6820(2) & $0$&034(30) & R &  $-0$&07(3)\\
B6863 & $0$&024(20) & R &  $-0$&06(2)\\
G029 & $0$&009(9) & L &  $-0$&068(9)\\
G018 & $0$&016(19) & L &  $-0$&061(19)\\[2pt]
\multicolumn{4}{r}{
{\footnotesize\it weighted mean:}} & $-0$&0679(9)\\
\hline\\[-8pt]
\end{tabular}
\end{table}

\section{Conclusions and future prospects}
\label{Sec6}

With the Effelsberg 100-m radio telescope, we carried out simultaneous observations
of Class~I methanol masers at frequencies 25, 36 and 44~GHz
towards 22 Galactic targets located at Galactocentric distances from 5 to 13 kpc. Comparison with
previous observations of the 44~GHz masers performed 6-10 years earlier with
the KVN towards the same targets confirmed the kinematic stability of the Class~I maser line profiles and
revealed a systematic shift of $0.013\pm0.005$~\kms\ in radial velocities between the
two telescopes.

Lankhaar \etal\ (2016, 2018) developed from first principles a quantum-chemical model for 
the hyperfine structure
in methanol and computed frequency shifts of individual HF components relative to the centres of
various torsion-rotational multiplets. Among others, they assumed that
in every multiplet there are only a few specific (`favored')
HF transitions (with largest Einstein A coefficients) which predominantly form masers. 

In our previous study (L22), by analyzing Class~I masers at 44 and 95~GHz in $A$-methanol
observed towards 19 Galactic targets, we confirmed the model predictions that each of
the $7_0-6_1A^+$ (44~GHz) and $8_0-7_1A^+$ (95~GHz) transitions has two `favored'
HF components~-- one to the left and one to the right of the multiplet centre~-- 
and found that only one of them is masering in any given source.

The present observations were carried out with the aim to test whether the assumption of 
the hyperfine-specific effects in the maser action is valid also for
other Class~I masers, in particular for transitions at 25 and 36~GHz in $E$-methanol,
and to identify the masering components in the corresponding HF multiplets.

The obtained main results are as follows:
\begin{enumerate}
\item
We re-processed the laboratory spectra of several transitions in the $E$-type ground
torsional state ($v_t = 0$) of CH$_3$OH
recorded in 2012 with the microwave molecular beam spectrometer of the Hannover University.
The S/N ratio in the obtained line profiles
made it possible to deconvolve the hyperfine structure of the multiplets,
in particular that of the $5_2-5_1E$ (24959 MHz) transition which is
usually observed as a strong Class~I maser. `Favored' HF components were identified.

\item
Among the observed 22 targets selected from methanol maser catalogs
compiled on base of the 44~GHz observations, only 9 targets
(10 maser sources) showed
the simultaneous activity of masers at 25, 36 and 44~GHz. A detailed analysis of all observed
maser line profiles was carried out and the line peak (radial) velocities were determined. 
With these velocities, the differences
$\Delta V_{\scriptscriptstyle 36-25}$ and $\Delta V_{\scriptscriptstyle 44-25}$ 
were calculated. At 25~GHz, the maser line
widths are much narrower than the splitting between the `favored' HF components, 
which confirms the result previously obtained
for the 44~ and 95~GHz masers: namely, that only one HF component is acting in any one maser source.

\item
As expected, the calculated values $\Delta V_{\scriptscriptstyle 44-25}$ form separate groups. 
With masering
HF components at 44~GHz known from the previous study and using the HF
component structure obtained from the laboratory spectrum of the $5_2-5_1E$ multiplet as a template,  we
compared the differences between the groups with the splitting between the `favored' components in the template and
determined for every source which particular HF component acted as the 25~GHz maser. The $5_2-5_1E$ multiplet
has three `favored' components. Among 10 sources we detected 6 masers formed by the low-frequency
component, 4 by the middle-frequency component, and none with the high-frequency component.

\item
The velocity differences $\Delta V_{\scriptscriptstyle 36-25}$ form separate groups as well.
The template of the HF structure at 36~GHz was computed with the quantum-chemical model
by Lankhaar \etal\ (2016, 2018). Using the same
procedure, i.e., comparing the differences between the groups with the splitting between the `favored' components
in the template, we identified the masering 
components in the 36~GHz masers. The model predicts 4 `favored'
components, but we observed as masers only three of them. 
Again no masers with the high-frequency component were found.

\item
The laboratory rest frequencies of the $4_{-1}-3_0E$ (36~GHz), $7_0-6_1A^+$ (44~GHz) 
and $8_0-7_1A^+$ (95~GHz) multiplet
centres were known with accuracies of about 10~kHz, whereas the rest frequency of the $5_2-5_1E$ (25~GHz) transition was
measured with an order of magnitude higher accuracy. 
Using the common procedure of the line alignment and accounting for
a particular HF component which forms a maser in a specific source, 
we corrected the rest frequencies of these transitions.
The refined rest frequency values are as follows:
$f_{\scriptscriptstyle 36} = (36169.2488\pm 0.0002_{\scriptscriptstyle\rm stat} \pm
0.0004_{\scriptscriptstyle\rm sys})$~MHz.
$f_{\scriptscriptstyle 44} = (44069.4176\pm0.0002_{\scriptscriptstyle\rm stat} \pm    
0.0004_{\scriptscriptstyle\rm sys})$~MHz,
and
$f_{\scriptscriptstyle 95} = (95169.4414\pm 0.0003_{\scriptscriptstyle\rm stat} \pm
0.0004_{\scriptscriptstyle\rm sys})$~MHz.

\end{enumerate}

The present analysis is based entirely on the kinematic characteristics of the maser line profiles. For
all masers both in $A$- and $E$-methanol, we confirm that they are formed by only one HF component and
this component varies from source to source. 
Among components indicated as `favored', we do not observe in the $E$-methanol
masers at 25 and 36~GHz the high-frequency HF component acting.
In general, it is quite obvious, that both the pumping of the inverted population and the formation
of the maser radiation itself should be in some way modulated by physical and chemical conditions
in the gain medium.  It is also clear that this modulation should be fine-tuned, 
taking into account very small frequency differences between particular HF components.
However, at the moment the mechanism behind the anisotropic pumping of methanol masers remains
completely obscure.
It is to note that anomalous effects related to the HF structures in different molecules are well known, 
but in spite of many studies (e.g., Field 1985; Park 2001; Camarata \etal\ 2015; Zhou \etal\ 2020;
Lankhaar \etal\ 2018, 2024) the problems are still far from being solved.

\section*{Acknowledgments}
We are grateful to the staff of the Effelsberg 100-m
telescope observatory for assistance in our observations.
We also thank J.-U. Grabow for laboratory measurements of methanol lines
at the Hannover University. 
S.A.L. is supported in part by the Russian Science Foundation
under grant No.~23-22-00124.
O.S.B. acknowledges financial support from the Italian Ministry of University and Research~-- 
Project Proposal CIR01-00010.

\section*{Data Availability}

The data underlying this article will be shared on reasonable request to the
corresponding author.

\bsp    
\label{lastpage}

\begin{thebibliography}{}
\bibitem{}
Argon A.L., Reid M.J., Menten K.M., 2000, ApJS, 129, 159
\bibitem{}
Bae J.-H., Kim K.-T., Youn S.-Y., Kim W.-J., Byun D.-Y., Kang H., Oh C. S., 
2011, ApJS, 196, 18
\bibitem{}
Barrett A. H., Schwartz P. R., Waters J. W., 1971, ApJ, 168, L101
\bibitem{}
Batrla W., Matthews h. E., Menten K. M., Walmsley C. M., 1987, Nature, 326, 49
\bibitem{}
Bayandina O. S., Burns R. A., Kurtz S. E., Shakhvorostova N. N., Val'tts I. E., 2019, ApJ, 884, 140
\bibitem{}
Belov S. P., \etal, 2016, J. Chem. Phys. 145, 024307
\bibitem{}
Breen S. L., Contreras Y., Dawson J. R., Ellingsen S. P., Voronkov M. A., McCarthy T. P., 
2019, MNRAS, 484, 5072
\bibitem{}
Coudert L. H., Guttl\'e C., Huet T. R., Grabow J.-U., Levshakov S. A., 2015, J. Chem. Phys., 143, 044304 
\bibitem{} 
Camarata M. A., Jackson J. M., Chambers E., 2015, ApJ, 806, 74
\bibitem{}
Cragg D. M., Sobolev A. M., Goddfrey P. D., 2005, MNRAS, 360, 533
\bibitem{}
Crutcher R.M., Kemball A.J., 2019, FrASS, 6, 66
\bibitem{}
Cyganowski C. J., Brogan C. L., Hunter T. R., Churchwell E., 2009, ApJ, 702, 1615
\bibitem{}
Dapr\`a M., \etal, 2017, MNRAS, 472, 4434
\bibitem{}
Dore L., \etal, 2004, A\&A, 413, 1177
\bibitem{}
Fontani F., Cesaroni R., Furuya R. S., 2010, A\&A, 517, A56
\bibitem{}
Field D., 1985, MNRAS, 217, 1
\bibitem{}
Grabow J.-U., 2004, ``Chemische Bindung und interne Dynamik in gro{\ss}en isolierten Molek\"ulen:
Rotationsspektroskopische Untersuchung'', Habilitationsschrift, Dem Fachbereich Chemie der Universit\"at Hannover
\bibitem{}
Grabow J.-U., 2011, ``Fourier Transform Microwave Spectroscopy Measurement and
Instrumentation'', Handbook of High-Resolution Spectroscopy, M. Quack and F. Merkt (eds.),
John Wiley \& Sons, Chichester, 723pp
\bibitem{}
Green J. A., \etal, 2017, MNRAS, 469, 1383
\bibitem{}
Haschick A. D., Menten K. M., Baan W. A., 1990, ApJ, 354, 556
\bibitem{}
Heuvel J., Dymanus A., 1973a, J. Mol. Spectrosc. 45, 282
\bibitem{}
Heuvel J., Dymanus A., 1973b, J. Mol. Spectrosc. 47, 363
\bibitem{}
Hougen J., Kleiner I., Godefroid M., 1994, J. Mol. Spectrosc., 163, 559
\bibitem{}
Jansen P., Xu L.-H., Kleiner I., Ubachs W., Bethlem H. L., 2011, Phys. Rev. Lett. 106, 100801
\bibitem{}
Kalenskij S. V., \etal, 1992, Proceed. Conf., Arlington, VA, Mar. 9-11, (A93-52776 23-90), pp.191-194
\bibitem{}
Kalenskii S. V., Slysh V. I., Val'tts I. E., Dzura A. M., 1996, IAUS, 170, 49
\bibitem{}
Kalenskii S. V., \etal, 2010, MNRAS, 405, 613
\bibitem{}
Kang J., Byun D.-Y., Kim K.-T., Kim J., Lyo A.-R., Vlemmings W. H. T., 2016, ApJS, 227, 18
\bibitem{}
Kim C.-H., Kim K.-T., Park Y.-S., 2018, ApJS, 236, 31 
\bibitem{}
Kim W.-J., Kim Kee-Tae., Kim Kwang-Tae, 2019, ApJS, 244, 33
\bibitem{}
Klein B., Hochg\"urtel S., Kr\"amer I., Bell, A., Meyer K., G\"usten R., 
2012, A\&A, 542, L3
\bibitem{}
Kurtz S., Hofner P., Alvarez C. V., 2004, ApJS, 155, 149
\bibitem{}
Ladeyschikov D. A., Bayandina O. S., Sobolev A. M., 2019, AJ, 158, 233
\bibitem{}
Ladeyschikov D. A., Urquhart J. S., Sobolev A. M., Breen S. L., Bayandina O. S., 2020, AJ, 160, 213
\bibitem{}
Lankhaar B., Groenenboom G.C., van der Avoird A., 2016, J. Chem. Phys., 145, 244301 
\bibitem{}
Lankhaar B., Vlemmings W., Surcis, G., van Langenvelde H.J., Groenenboom G.C.,
van der Avoird A., 2018, NatAs, 2, 145L 
\bibitem{}
Lankhaar B., Surcis G., Vlemmings W., Impellizzeri V., 2024, A\&A, 683, 117
\bibitem{}
Leurini S., Menten K. M., Walmsley C. M., 2016, A\&A, 592, A31
\bibitem{}
Levshakov S. A., Kozlov M. G., Reimers D., 2011, ApJ 738, 26
\bibitem{}
Levshakov S. A., \etal, 2019, MNRAS, 487, 5175
\bibitem{}
Levshakov S. A., \etal, 2022, MNRAS, 511, 413 [L22]
\bibitem{}
Liechti S., Wilson T. L., 1996, A\&A, 314, 615
\bibitem{}
Mehrotra S. C., Dreizler H., Mäder H., 1985, Z. Naturforschung, 40a, 683
\bibitem{}
Menten K. M., Walmsley C. M., Henkel C., Wilson T. L., 1986, A\&A, 157, 318
\bibitem{}
Menten K. M., Walmsley C. M., Henkel C., Wilson T. L., 1988, A\&A, 198, 267
\bibitem{}
Menten K. M., 1991, in Proc. Third Haystack observatory Meeting, eds.
A. D. Haschick \& P. T. P. Ho (San Francisco: ASP), p.~119
\bibitem{}
Menten K. M., Reid M. J., Pratar P., Moran J. M., Wilson T. L., 1992, ApJ, 401, L39
\bibitem{}
Minier V., Conway J. E., Booth R. S., 2001, A\&A, 369, 278
\bibitem{}
Momjian E., Sarma A. P., 2019, ApJ, 872, 12
\bibitem{}
M\"uller H. S. P., Menten K. M., M\"ader H., 2004, A\&A, 428, 1019
\bibitem{}
Pagani L., Daniel F., Dubernet M.-L., 2009, A\&A, 494, 719
\bibitem{}
Pagani L., \etal, 2017, A\&A, 604, 32
\bibitem{}
Park Y.-S., 2001, A\&A, 376, 348
\bibitem{}
Pickett H. M., Poynter R. L., Cohen E. A., Delitsky M. L., Pearson J. C., M\"uller H. S. P., 1998, JQSRT, 60, 883
\bibitem{}
Plambeck R. L., Menten K. M., 1990, ApJ, 364, 555
\bibitem{}
Sarma A. P., Momjian E., 2009, ApJ, 705, L176
\bibitem{}
Sarma A. P., Momjian E., 2020, ApJ, 890, 6
\bibitem{}
Sobolev A. M., Deguchi S., 1994, A\&A, 291, 569
\bibitem{}
Tsunekawa S., Ukai T., Toyama A., Takagi K., 1995,
Toyama Microwave Atlas for spectroscopists and astronomers.
Department of Physics, Toyama University, Japan,
Available at: https://www.sci.u-toyama.ac.jp/phys/4ken/atlas/
\bibitem{}
van Terwisga S. E., Hacar A., van Dishoeck E. F., 2019, A\&A, 628, A85
\bibitem{}
Vlemmings W. H. T., 2008, A\&A, 484, 773
\bibitem{}
Voronkov M. A., Caswell J. L., Ellingsen S. P., Green J. A., Breen S. L., 2014, MNRAS, 439, 2584
\bibitem{}
Vorotyntseva J. S., Kozlov M. G., Levshakov S. A., 2024, MNRAS, 527, 2750
\bibitem{}
Vorotyntseva J. S., Levshakov S. A., 2024, AApTr, 34, No.2 (in press),
arXiv:2310.08231 [physics.atom-ph]
\bibitem{}
Wenner N., Sarma A. P., Momjian P., 2022, ApJ, 930, 114
\bibitem{}
Xu L.-H., \etal, 2008, J. Mol. Spectrosc., 251, 305
\bibitem{}
Yang W., \etal, 2020, ApJS, 248, 18
\bibitem{}
Zhou D.-D., \etal, 2020, A\&A, 640, 114 
\end{thebibliography}
\end{document}